\documentclass[twocolumn,trackchanges]{aastex631}

\usepackage{CJK}

\newcommand{\petit}{\texttt{petitRADTRANS}}

\newcommand{\teff}{T$_{\rm eff}$}

\newcommand{\kms}{$\rm km\,s^{-1}$}
\newcommand{\kp}{$K_{\rm p}$}
\newcommand{\dvsys}{$\Delta v_{\rm sys}$}
\newcommand{\kpvsys}{$K_{\rm p}-\Delta v_{\rm sys}$}
\newcommand{\hd}{HD~209458}
\newcommand{\hdb}{HD~209458~b}
\newcommand{\caltech}{Department of Astronomy, California Institute of Technology, Pasadena, CA 91125, USA}
\newcommand{\gps}{Division of Geological \& Planetary Sciences, California Institute of Technology, Pasadena, CA 91125, USA}
\newcommand{\ucsc}{Department of Astronomy \& Astrophysics, University of California, Santa Cruz, CA95064, USA}
\newcommand{\keck}{W. M. Keck Observatory, 65-1120 Mamalahoa Hwy, Kamuela, HI 96743, USA}
\newcommand{\ucla}{Department of Physics \& Astronomy, 430 Portola Plaza, University of California, Los Angeles, CA 90095, USA}
\newcommand{\jpl}{Jet Propulsion Laboratory, California Institute of Technology, 4800 Oak Grove Dr.,Pasadena, CA 91109, USA}
\newcommand{\ucsd}{Department of Astronomy \& Astrophysics, University of California San Diego, La Jolla, CA 92093, USA}

\newcommand{\northwestern}{Center for Interdisciplinary Exploration and Research in Astrophysics (CIERA) and Department of Physics and Astronomy,
Northwestern University, Evanston, IL 60208, USA}
\newcommand{\arizona}{James C. Wyant College of Optical Sciences, University of Arizona,
Meinel Building 1630 E. University Blvd., Tucson, AZ 85721, USA}

\received{\today}

\shorttitle{Joint KL retrieval of \hdb}
\shortauthors{Finnerty et al.}

\graphicspath{{./}{}}

\begin{document}
\begin{CJK*}{UTF8}{gbsn}

\title{The watery atmosphere of HD~209458~b revealed by joint $K$- and $L$-band high-resolution spectroscopy}

\correspondingauthor{Luke Finnerty}
\email{lfinnert@umich.edu}

\author[0000-0002-1392-0768]{Luke Finnerty}
\affiliation{\ucla}

\author{Julie Inglis}
\affiliation{\caltech}

\author[0000-0002-0176-8973]{Michael P. Fitzgerald}
\affiliation{\ucla}

\author[0000-0002-1583-2040]{Daniel Echeverri}
\affiliation{\caltech}

\author[0000-0001-5213-6207]{Nemanja Jovanovic}
\affiliation{\caltech}

\author{Dimitri Mawet}
\affiliation{\caltech}
\affiliation{\jpl}

\author{Geoffrey A. Blake}
\affiliation{\gps}

\author[0000-0002-6525-7013]{Ashley Baker}
\affiliation{\caltech}

\author{Randall Bartos}
\affiliation{\jpl}

\author[0000-0003-4737-5486]{Benjamin Calvin}
\affiliation{\caltech}
\affiliation{\ucla}

\author{Sylvain Cetre}
\affiliation{\keck}

\author[0000-0001-8953-1008]{Jacques-Robert Delorme}
\affiliation{\keck}
\affiliation{\caltech}

\author{Greg Doppmann}
\affiliation{\keck}

\author[0000-0001-9708-8667]{Katelyn Horstman}
\affiliation{\caltech}
\altaffiliation{NSF Graduate Research Fellow}

\author[0000-0002-5370-7494]{Chih-Chun Hsu}
\affiliation{\northwestern}

\author[0000-0002-4934-3042]{Joshua Liberman}
\affiliation{\caltech}
\affiliation{\arizona}

\author[0000-0002-2019-4995]{Ronald A. L\'opez}
\affiliation{\ucla}

\author{Evan Morris}
\affiliation{\ucsc}

\author{Jacklyn Pezzato-Rovner}
\affiliation{\caltech}

\author[0000-0003-2233-4821]{Jean-Baptiste Ruffio}
\affiliation{\ucsd}

\author[0000-0003-1399-3593]{Ben Sappey}
\affiliation{\ucsd}

\author{Tobias Schofield}
\affiliation{\caltech}

\author{Andrew Skemer}
\affiliation{\ucsc}

\author[0000-0001-5299-6899]{J. Kent Wallace}
\affiliation{\jpl}

\author[0000-0003-0354-0187]{Nicole L. Wallack}
\affiliation{Earth and Planets Laboratory, Carnegie Institution for Science, Washington, DC 20015, USA}

\author[0000-0003-0774-6502]{Jason J. Wang (王劲飞)}
\affiliation{\northwestern}

\author[0000-0002-4361-8885]{Ji Wang (王吉)}
\affiliation{Department of Astronomy, The Ohio State University, 100 W 18th Ave, Columbus, OH 43210 USA}

\author[0000-0002-6171-9081]{Yinzi Xin}
\affiliation{\caltech}

\author[0000-0002-6618-1137]{Jerry W. Xuan}
\affiliation{\caltech}

\begin{abstract}

We present a joint analysis of high-resolution $K$- and $L$-band observations of the benchmark hot Jupiter \hdb\ from the Keck Planet Imager and Characterizer (KPIC). One half night of observations were obtained in each bandpass covering similar pre-eclipse phases. The two epochs were then jointly analyzed using our atmospheric retrieval pipeline based on \petit\ to constrain the atmospheric pressure-temperature profile and chemical composition. Consistent with recent results from \textit{JWST} observations at lower spectral resolution, we obtain an oxygen-rich composition for  \hdb\ ($\rm C/O < 10^{-3}$ at 95\% confidence) and a lower limit on the volatile metallicity similar to the solar value  ($\rm [(C+O)/H] > -0.2$ at 95\% confidence). Leveraging the large spectral grasp of the multi-band observations, we constrain the H$_2$O mixing ratio to $\rm \log H_2O_{VMR} > -3.1$ at 95\% confidence, and obtain 95\% upper limits on the atmospheric mixing ratios of CO ($<10^{-4.8}$), CH$_4$ ($<10^{-4.5}$), NH$_3$ ($<10^{-5.8}$), H$_2$S ($<10^{-3.3}$), and HCN ($<10^{-5.6}$). The limits on CH$_4$, NH$_3$, and HCN are consistent with recent results from \textit{JWST} transmission spectroscopy, demonstrating the value of multi-band ground-based high resolution spectroscopy for precisely constraining trace species abundances in exoplanet atmospheres. The retrieved low-C/O, moderate-metallicity composition for \hdb\ is consistent with formation scenarios involving late accretion of substantial quantities of oxygen-rich refractory solids and/or ices. 

\end{abstract}


\section{Introduction} \label{sec:intro}

\hdb\ was the first exoplanet to be detected via the transit technique \citep{henry2000}. Its detection established that hot Jupiters are gas giants, the first example of exoplanet characterization. From this first discovery has followed a long history of characterization efforts, including the first atmospheric detection of an exoplanet via high-resolution cross-correlation spectroscopy \citep[HRCCS,][]{snellen2010}. Recently, \hdb\ has been observed in transmission with \textit{JWST}, revealing strong CO$_2$ and H$_2$O absorption features and a modestly metal-enriched, low-C/O ratio bulk composition \citep{xue2024}. Recent high-resolution transmission spectroscopy from CRIRES+ has independently confirmed the low C/O ratio, moderately metal-enriched composition \citep{blain2024hd209}. These analyses have also ruled out significant quantities of  C$_{2}$H$_{2}$ and CH$_{4}$, which were previously reported in \hdb\ by \citet{giacobbe2021}, as well as HCN, which was previously reported by both \citet{giacobbe2021} and \citet{hawker2018}. 

\hdb\ was observed as part of the KPIC $K$-band (2.1--2.5 $\mu$m) hot Jupiter survey, and again with KPIC in the $L$ band (3.0--4.1 $\mu$m). These two bandpasses span strong spectral features of CO, H$_2$O, CH$_4$, NH$_3$, H$_2$S, and HCN, presenting an opportunity to obtain a comprehensive inventory of the expected volatile species in hot/warm Jupiter atmospheres and place constraints not just on the atmospheric C/O ratio, but also on S/O and N/O. Recent detections of H$_2$S \citep{fu2024, inglis2024} and SO$_2$ \citep{rustamkulov2023, powell2024, dyrek2024} have highlighted the importance of sulfur species as tracers of both photochemistry \citep{tsai2023} and the planet formation process \citep{turrini2021, crossfield2023}. While clear detections of nitrogen species have remained elusive, models show that combining C/O and N/O ratios can break degeneracies in the predicted formation location of gas giants \citep{cridland2020}. 

Additionally, the use of wide bandpasses appears to significantly improve the accuracy of HRCCS retrievals of atmospheric thermal structure. Retrievals using KPIC $K$-band data alone have often preferred pressure-temperature ($P-T$) profiles running substantially colder than expected \citep[e.g.][]{finnerty2025b}, typically counteracted by a scaling parameter $>1$. In contrast, retrievals from IGRINS, which covers the full $H$ and $K$ bands, have found $P-T$ profiles more in line with expectations \citep[e.g.][]{line2021}. This may be a result of the loss of continuum during the data processing for HRCCS analysis. For typical hot Jupiter temperatures, the $K$-band $F_p/F_s$ changes relatively little with wavelength, and between the multiplicative scaling factor and the $P-T$ parameters there is sufficient degeneracy in retrieved parameters that errors in the absolute temperatures can be countered by changes in other parameters. Adding $H$ or $L$ band data introduces a much larger variation in $F_p/F_s$ with wavelength, requiring an accurate temperature in order to match $F_p/F_s$ over the entire range of observations. 

Section \ref{sec:obs} describes the observations, data reduction procedures, and atmospheric retrieval setup. Section \ref{sec:res} presents the results of our atmospheric retrievals. Section \ref{sec:disc} discusses the retrieved composition in comparison to other studies of \hdb. Section \ref{sec:conc} concludes.

 \begin{deluxetable}{ccc}
    \tablehead{\colhead{Property} & \colhead{Value} & \colhead{Ref.}}
    \startdata
        & \textbf{\hd} & \\
        \hline
        RA & 22:03:10.81 &  \citet{gaiaedr3} \\
        Dec & +18:53:03 &  \citet{gaiaedr3} \\
        Sp. Type & G0V &  \\
        $K_{\rm mag}$ & $6.31\pm0.03$ & \citet{cutri2003} \\
        Mass & $1.07\pm0.05\rm\ M_\odot$ & \citet{rosenthal2021}  \\
        Radius & $1.155\pm0.02\rm\ R_\odot$ & \citet{bonomo2017} \\
        \teff & $6000\pm100$ K & \citet{rosenthal2021} \\
        $\log g$ [cgs] & $4.31\pm0.03$ & \citet{rosenthal2021}  \\
        $v\sin i$ & $4.5\pm0.5$ \kms & \citet{bonomo2017} \\
        $v_{\rm rad}$ & $-15.0\pm0.3$ \kms & \citet{gaiaedr3}  \\
        $\rm [Fe/H]$ & $0.0\pm0.06$ & \citet{rosenthal2021} \\
        \smallskip \\
        \hline
         & \textbf{\hdb} & \\
        \hline
        Period & $3.5247496\pm3\times10^{-7}$  days & \citet{kokori23} \\
        $\rm T_{\rm trans}$ & $\rm JD\ 2455420.8446\pm2\times10^{-4} $ & \citet{kokori23}   \\
        $a$ [AU] & $0.0463\pm0.0007$ & \citet{rosenthal2021}\\
        $i$ [deg] & $86^\circ.71\pm0^\circ.05$ & \citet{kokori23} \\
        $K_{\rm p}$ & 147 \kms  & Est. \\
        Mass & $0.68\pm0.02\rm\ M_J$ & \citet{bonomo2017}\\
        Radius & $1.36\pm0.02 \rm\ R_J$ & \citet{bonomo2017} \\
        $\rm T_{\rm eq}$ &  1450 K & Est. \\
        C/O & $<10^{-3}$ & This work \\
        $\rm [C/H]$ & $<-0.9$ & This work \\
        $\rm [O/H]$ & $>0.0$ & This work \\
    \enddata
    \caption{Stellar and planetary properties for the \hd\ system. Limits from this work are given at 95\% confidence, and abundances relative to solar assume the \citet{asplund2021} solar abundances.} 
    \label{tab:props}
\end{deluxetable}

\section{Observations and Data Reduction}\label{sec:obs}

\subsection{Observations}

\hd\ was observed with Keck II/KPIC \citep{nirspec, nirspecupgrade, nirspecupgrade2, kpic, echeverri2022, kpicII} on UT 29 June 2024 in $K$ and on 21 August 2024 in $L$. Each epoch is summarized below. 

\subsubsection{\textit{K}-band}

$K$ band observations were obtained on UT 2024 June 29 from 10:52 to 15:18, covering a phase range of 0.431--0.483 and nominal planetary radial velocity ranging from 62 \kms\ to 16 \kms. We used a 120 second exposure time in order to stay under the detector non-linearity threshold, obtaining 105 frames on KPIC science fiber 4. Similar to previous KPIC $K$-band observations \citep{finnerty2025a, finnerty2025b}, daytime calibration frames were used for background correction. The secondary eclipse is expected to begin at phase $\phi = 0.479$, and we therefore omit the last 8 frames from subsequent analysis. 

Conditions were good through most of the observations, aside from a brief period (approximately five minutes) of poor AO correction as \hd\ transited zenith. Instrument throughput from the top of the atmosphere was $\sim3-4\%$. Observations began as \hd\ was rising, at an airmass of 1.36, decreasing to 1.0 at transit and increasing to 1.06 by the end of the sequence. 

\subsubsection{\textit{L} band}

$L$-band observations were obtained on UT 2024 August 21 from 6:05 to 10:25, covering a phase range of 0.412--0.462 and a nominal planetary radial velocity range from 77 \kms\ to 35 \kms. We used 60 second exposures to avoid non-linearity from thermal background, and used an ABBA nodding pattern between science fibers 2 and 4 in order to account for the large and variable background. A total of 90 frames were obtained for each fiber.  

Conditions were poor throughout the observations, with $\sim2''$ typical seeing and brief periods where AO correction failed entirely. While we experimented with omitting up to 18 frames with low throughput, we found this did not significantly impact the retrievals, the present results based on the full time series. Observations began as \hd\ was rising at an airmass of 2.06, and ended at an airmass of 1.0. 

\subsection{Data Reduction}

Both epochs were reduced using the modified KPIC DRP\footnote{\href{https://github.com/kpicteam/kpic_pipeline/}{https://github.com/kpicteam/kpic\_pipeline/}} described in \citet{finnerty2025a, finnerty2025b}. For the $K$-band data, background subtraction was performed using afternoon calibration frames, while for the $L$-band data we performed an AB subtraction to account for variations in the slit background.

We use the Gaussian-Hermite trace profile/line-spread function (LSF) fitting procedure previously described in \citet{finnerty2025b} to determine profile weights for optimal extraction and the instrumental LSF. The trace profile is the product of a Gaussian profile and a weighted sum of the first five Hermite polynomials, which can capture asymmetries in the profile.  The fitting procedure first stacks the entire science data sequence to maximize the SNR. For each wavelength channel, we then fit a six-parameter Gaussian-Hermite model, with free parameters for the Gaussian width, an offset from the nominal trace center, and the weights of the first four non-trivial Hermite polynomials. The resulting fit parameters for each order are then smoothed with a 51-pixel Gaussian filter to determine the values used for the final extraction/LSF weights. This smoothing rejects outliers and reduces the impact of frame-to-frame fitting errors, and results in variations in the trace profile on a similar spatial scale to the NIRSPEC resolution variations reported in \citet{nirspecupgrade2}. For the LSF, we then stretch the trace profiles by a factor of 1.14, based on known asymmetries in the NIRSPEC optics and measurements reported in \citet{Finnerty2022}.

For wavelength calibration, we observed the red giants HIP 95771 ($K$-band) and HIP 81497 ($L$-band) prior to the observations of \hd. The $K$-band wavelength solution was fit using the KPIC DRP and manually checked against a stellar $\times$ telluric model to verify the accuracy of the wavelength solution in the 3 orders from 2.09--2.27 $\mu$m, which have relatively few strong lines for calibration.  The included $K$-band data cover roughly $2.1-2.5\ \mu\rm m$, with significant gaps. The wavelength coverage is shaded in Figure \ref{fig:specplot}.  The median FWHM of the LSF in the $K$-band corresponds to a spectral resolution $R = \lambda/\Delta\lambda\approx38,000$ for the bluest two orders, and declines steadily to $\sim 34,000$ by the reddest order.

For the $L$-band wavelength solution, we first obtained an initial guess at the wavelength solution from the NIRSPEC Echelle Format Simulator, which was used to manually tune the coefficients of a 2nd-order polynomial for each order by comparing the observed spectrum to the stellar $\times$ telluric model. This was then input as the initial wavelength solution for the KPIC DRP scripts. However, the DRP scripts failed to produce a good solution for the three reddest orders (3.61--4.06 $\mu$m), which have relatively weak spectral features and relatively strong fringing effects. For these orders, we use a wavelength solution based on manually tuning polynomial coefficients to obtain a good match between the model telluric/stellar absorption features and the data. The included $L$-band data span roughly $3.05-4.05\ \mu\rm m$, but with significantly larger gaps compared to the $K$-band data, as can be seen in Figure \ref{fig:specplot}. The LSF measurements indicate a substantially lower spectral resolution compared to the $K$ band, with $R\sim28,000$ in the bluest order declining to $R\sim20,000$ at 4 $\mu$m.

\subsection{Detrending}

High-resolution spectroscopic time series include time-varying telluric and instrumental effects significantly stronger than the planet signal \citep[e.g. Figure 1 of][]{finnerty2024}, necessitating a detrending procedure to enable planet detection. This procedure is applied to the spectral time series for each order individually.

We first divide each individual spectrum by its median, establishing a common scale for the time series. We then compute the median of the time series and mask wavelength channels with telluric transmission $<70\%$, and the 3\% of remaining channels with the highest temporal variance. This masks regions strongly impacted by tellurics and/or bad pixels. We then divide the entire time series by the median spectrum, removing static stellar and telluric features. Finally, we clip the first and last 50 wavelength channels of each order, where the instrumental blaze function leads to low fluxes. Values $>$6$\times$ the Median Absolute Deviation (MAD) are then clipped, removing any uncorrected bad pixels.

While this effectively removes static non-planetary features from the time series, time-varying telluric or fringing effects remain. While the fringing previously discussed in \citet{Finnerty2022, finnerty2023, finnerty2024, horstman2024} was largely resolved by the KPIC phase II upgrade \citep{kpicII}, tellurics are still sufficient to require further detrending. This is accomplished using principal component analysis (PCA) along the temporal axis of each time series. We performed separate retrievals omitting 4, 6, and 8 principal components. The dropped components are saved and added to the forward model of the planetary atmosphere, similar to the approach described in \citet{line2021}. The PCA is then repeated on the forward model with the added dropped components, which replicates any distortions in the planet model caused by PCA. This approach was previously used successfully in \citet{finnerty2024, finnerty2025a, finnerty2025b}.  

\subsection{Atmospheric Retrieval}

We use the retrieval pipeline described in \citet{finnerty2023, finnerty2024, finnerty2025a, finnerty2025b}. The pipeline uses \petit\ for radiative transfer \citep{prt:2019, prt:2020, Nasedkin2024}, and performs sampling using the \texttt{PyMultiNest} \citep{buchner2014} wrapper for \texttt{MultiNest} \citep{feroz2008, feroz2009, feroz2019}. A separate log-likelihood is calculated for each fiber on each night using the \citet{brogi2019} formalism, which are then summed for the sampling. We used 1600 live points and a convergence criteria of $\Delta \log z < 0.01$, similar to our previous retrievals. 

We continue to use the \citet{guillot2010} parameterization for the Pressure-Temperature ($P-T$) profile employed in our previous retrieval work \citep{finnerty2023, finnerty2024, finnerty2025a, finnerty2025b}, which consists of the log of the infrared opacity ($\log \kappa$), the log of the infrared/optical opacity ratio ($\log \gamma$), an intrinsic temperature ($\rm T_{int}$), and an equilibrium temperature ($\rm T_{equ}$). Previous analyses of emission observations have differed on the presence of a thermal inversion in \hdb. For our fiducial retrievals, we restrict $\log \gamma < 0$ as a prior, disallowing thermal inversions, but we relax this prior in Section \ref{sec:disc}.

In contrast with \citet{finnerty2025a, finnerty2025b}, we do not fit for $\log g$ as a free parameter. The mass and radius of \hdb\ are both well-constrained, and initial retrievals including $\log g$ found a strong degeneracy between $\log g$ and $\log \kappa$, resulting in no useful constraint on $\log g$. Similarly, we experimented with retrievals including a gray cloud opacity with the cloud opacity, cloud pressure, and $f_{SED}$ as free parameters, but found the cloud parameters were unconstrained. 

We fit for the planetary radial velocity semi-amplitude, \kp, the planet systemic radial velocity offset, \dvsys, and the rotational broadening kernel, $v_{\rm rot}$. In principle, systematic errors in the wavelength calibration between bandpasses could lead to a difference in \dvsys\ between the $K$ and $L$ band data. We address this by performing individual retrievals for the $K$ and $L$ band data sets in addition to the combined retrievals.

In order to minimize the size of the opacity tables loaded into memory, separate radiative transfer objects are instantiated for the $K$ and $L$ bands. Each is instantiated identically except for a difference in wavelength boundaries. Further performance improvements could be achieved by loading only opacities with significant opacity in a given band, e.g. not loading CO opacity for the $L$-band radiative transfer object, but this has not yet been implemented.

We fit for mass-mixing ratios of H$_2$O, CO, CH$_4$, NH$_3$, H$_2$S, $^{13}$CO, and H$_2$. All abundances are fixed with pressure. For H$_2$O, we used the opacities based on the \citet{polyansky2018} POKAZATEL linelist. For $^{12}$CO, we used the high-temperature opacities described in \citet{finnerty2023} generated from \texttt{exocross} \citep{exocross2018} based on the HITEMP linelist \citep{hitemp2020}, while for $^{13}$CO we use a lower-temperature opacity table based on \citet{rothman2013}. We use the \citet{hargreaves2020} linelist for CH$_4$, and the \citet{yurchenko2011} linelist for NH$_3$. H$_2$S opacity is based on the \citet{azzam2016} linelist, and HCN uses \citet{harris2006}. H$_2$ opacity is from \citet{rothman2013}. We also include $\rm H_2 - H_2$ CIA opacity from \citet{borysow2001, borysow2002} and $\rm H_2-He$ opacity from \citet{borysow1988, borysow1989a, borysow1989b}. 

For \hd, we use a PHOENIX model \citep{phoenix} with \teff$=6100$ K, $\log g = 4.5$, and solar metallicity, which we rotationally broaden to $v\sin i = 4.5$ \kms\ using the \citep{carvalho2023} algorithm. While we experimented with Doppler shifting the spectrum based on the stellar reflex motion and the frame-specific barycentric velocity, we found this had no significant impact on the final posterior but incurred a significant computational cost. We therefore Doppler shift the stellar template by only the nominal systemic velocity minus the median barycentric velocity, which is constant for all frames. This effectively results in a featureless/smoothed stellar model in the forward model, similar to the approach taken in \citet{line2021}. 

The \citet{brogi2019} log-likelihood function includes a multiplicative scaling factor applied to the forward model. Physically, this scaling can account for mismatches between the assumed and actual planetary/stellar radius ratio, $\rm H^-$ opacity, or other effects which lead to an achromatic change in the planet flux relative to the host star. However, over narrow bandpasses, this scaling factor can become degenerate with changes in the $P-T$ profile, which can produce a similar overall scaling of the strength of planet features compared to the host star. Following \citet{finnerty2025b}, we address this by adopting a log-normal prior on the scale factor with a mean of 1.

\section{Results}\label{sec:res}

\begin{figure*}
    \centering
    \includegraphics[width=0.3\linewidth]{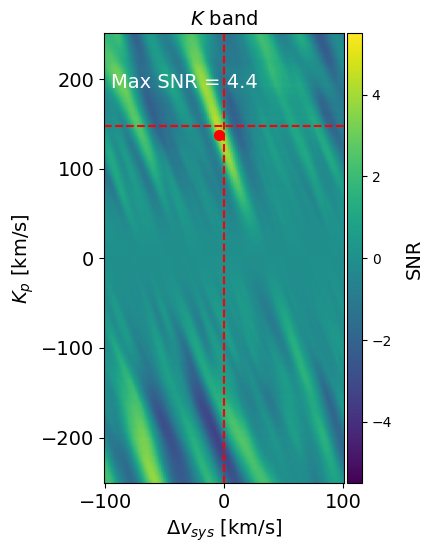}
    \includegraphics[width=0.3\linewidth]{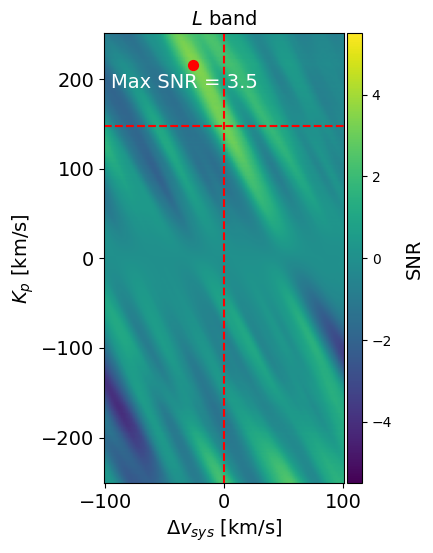}
    \includegraphics[width=0.3\linewidth]{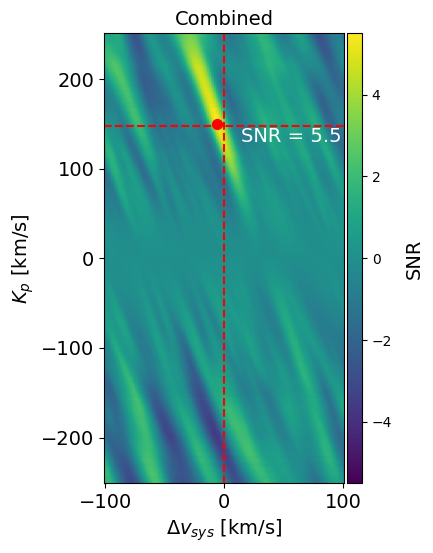}
    \caption{\kpvsys\ plots for the $K$-band (left), $L$-band (center) and combined (right) data sets using the maximum-likelihood retrieved planet model. \hdb\ is weakly detected ($\rm SNR\approx4$) in each band individually, improving to $SNR\sim5$ when both epochs are combined. The SNR is estimated by first median-subtracting each row of the \kpvsys\ plots, then dividing by the standard deviation of the negative $K_p$ region, which results in a slight offset of the maximum (indicated by a red dot) from the values in Table \ref{tab:priors}. See \citet{finnerty2024, finnerty2025b} for a more detailed discussion of the issues with estimating noise in $K_p - \Delta v_{sys}$ plots. }
    \label{fig:kpvsys}
\end{figure*}

\begin{figure*}
    \centering
    \includegraphics[width=0.3\linewidth]{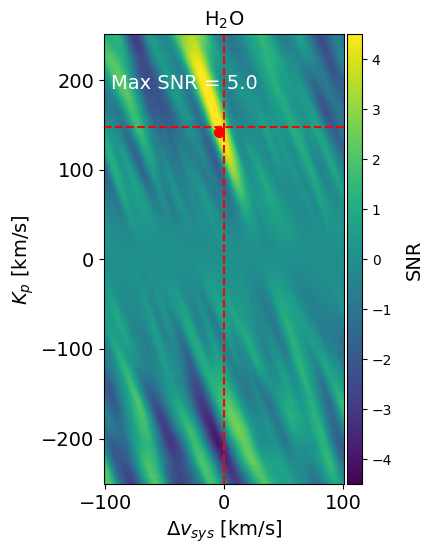}
    \includegraphics[width=0.3\linewidth]{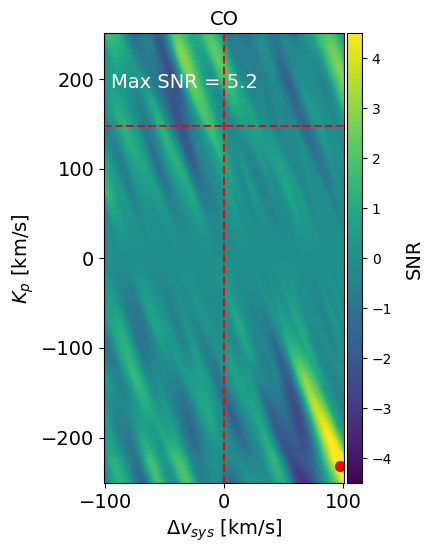}
    \includegraphics[width=0.3\linewidth]{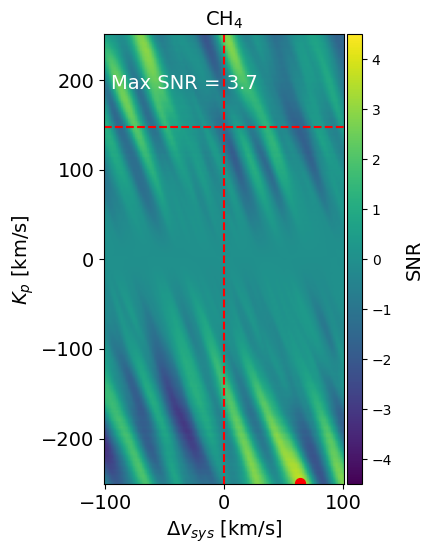}
    \includegraphics[width=0.3\linewidth]{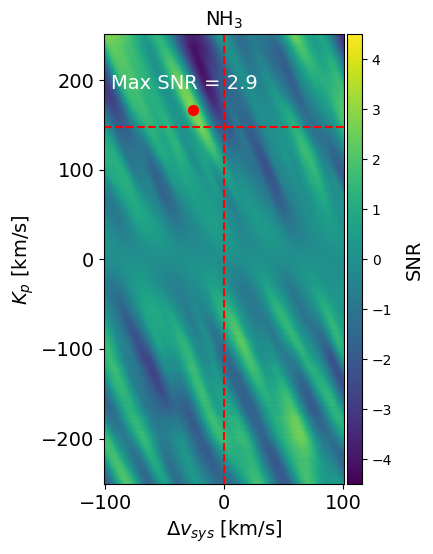}
    \includegraphics[width=0.3\linewidth]{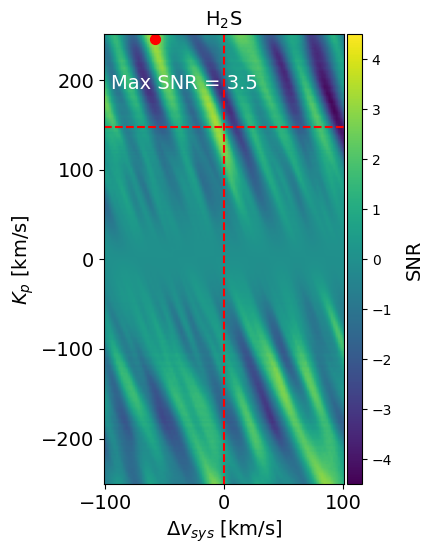}
    \includegraphics[width=0.3\linewidth]{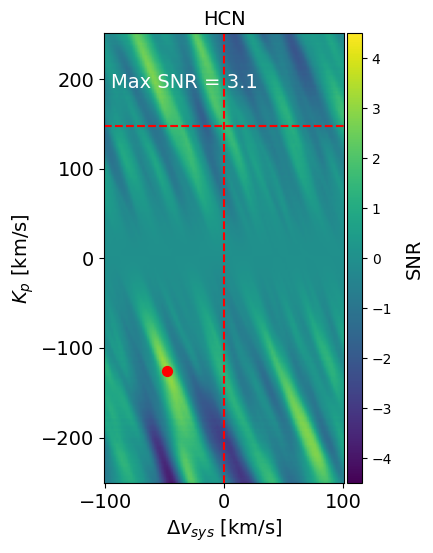}
    \caption{\kpvsys\ plots for each molecule. Planet models were generated using the maximum-likelihood parameters, setting the specified mass fraction to $10^{-1.5}$ and all others to $10^{-15}$. Only H$_2$O shows a clear peak consistent with the ephemeris of \hdb. While CO, H$_2$S, and HCN show $\rm SNR\sim3$ peaks consistent with the planet, comparison to other molecules shows that features of comparable strength regularly appear at random location in the \kpvsys\ space. CO shows a strong peak at very negative \kp, which may be a result of stellar or telluric contamination. }
    \label{fig:kpvsysmols}
\end{figure*}

Priors, maximum-likelihood parameters, and retrieved medians with $\pm34$\% confidence intervals from the 6 component joint retrieval are presented in Table \ref{tab:priors}. Our free-chemistry approach does not directly fit for atomic abundances or C/O, but we use the retrieved posteriors to construct posteriors for these parameters, and report the corresponding values in Table \ref{tab:priors} as derived parameters. The medians and confidence intervals for the 8 component case did not significantly differ from the values reported in Table \ref{tab:priors}, and we discuss the 4 component case below. The full corner plots for all retrievals are included in Appendix \ref{app:corner}.  We also present \kpvsys\ plots for each epoch and the combined data set in Figure \ref{fig:kpvsys}, demonstrating that \hdb\ is detected in the $K$ band data alone, weakly evident in the $L$ band data alone, and detected at significantly greater strength when combining the data from both bandpasses. Figure \ref{fig:kpvsysmols} presents the \kpvsys\ for individual molecules, demonstrating that the detection is dominated by H$_2$O, with no significant detection of any other species. 

\begin{deluxetable*}{ccccc}
    \tablehead{\colhead{Name} & Symbol  & \colhead{Prior} & \colhead{Retrieved Max-L} & \colhead{Retrieved Median}}
    \startdata
        log infrared opacity [$\rm cm^{2} g^{-1}$] & $\log \kappa$ &  Uniform($-4, 2$) &   $-1.1$ & $-0.9^{+0.6}_{-0.7}$ \\  
        log infrared/optical opacity & $\log \gamma$  &  Uniform($-3, 3$) & $-1.0$ & $-1.3^{+0.5}_{-0.4}$ \\
        Intrinsic temperature [K] & $\rm T_{int}$ & Uniform($10,500$) & $320$ & $170^{+130}_{-100}*$ \\
        Equilibrium temperature [K] & $\rm T_{equ}$ & Uniform($400,3000$) &  $1730$ & $1590^{+200}_{-240}$ \\ 
        $K_{\rm p}$ offset [\kms] & $\Delta K_{\rm p}$  & Uniform($-50, 50$) & $-20.0$ & $-23.0^{+13.0}_{-13.0}$ \\
        $v_{\rm sys}$ offset [\kms] & $\Delta v_{\rm sys}$ & Uniform($-15,15$) & $-0.6$ & $0.4^{+3.1}_{-3.1}$ \\
        Rotational velocity [\kms] & $v_{\rm rot}$ & Uniform($0,15$) & $11.5$ & $10.7^{+1.5}_{-1.6}$ \\
        log H$_2$O mass-mixing ratio & log H$_2$O  & Uniform($-12, -0.3$) & $-1.1$ & $-1.1^{+0.5}_{-0.7}\ (>-2.2)$ \\  
        log CO mass-mixing ratio & log CO & Uniform($-12,-0.3$) &  $-7.6$ & $-7.6^{+2.8}_{-2.8}\ (<-3.7)$ \\
        log CH$_4$ mass-mixing ratio & log CH$_4$ & Uniform($-12, -1$) &  $-7.0$ & $-7.7^{+2.8}_{-2.7}\ (<-3.9)$ \\
        log NH$_3$ mass-mixing ratio & log NH$_3$ & Uniform($-12, -1$) & $-7.5$ & $-8.3^{+2.3}_{-2.3}\ (<-4.9)$ \\  
        log H$_2$S mass-mixing ratio & log H$_2$S & Uniform($-12, -1$) & $-2.2$ & $-6.7^{+3.4}_{-3.4}\ (<-2.1)$ \\
        log HCN mass-mixing ratio & log HCN & Uniform($-12, -1$) & $-6.7$ & $-8.1^{+2.6}_{-2.5}\ (<-4.5)$ \\
        CO isotopologue ratio &   $\log \rm \frac{^{13}CO}{^{12}CO}$ & Uniform($-8, -0.5$) & $-3.6$ & $-4.2^{+2.3}_{-2.4}*$ \\
        log H$_2$ mass-mixing ratio & $\log \rm H_2$ & Uniform($-0.4,-0.05$) &  $-0.4$ & $-0.2^{+0.1}_{-0.1*}$ \\
        Scale factor & scale & LogNormal($0,0.1$) & $0.19$ & $0.02^{+0.09}_{-0.09}$ \\
        \smallskip \\
         & & \textbf{Derived Parameters} & \\
        \hline
        Carbon/oxygen ratio & C/O & - & $3\times10^{-6}$ & $<10^{-3}$ \\
        Carbon abundance & [C/H] & - & $-4.0$ & $<-0.9$ \\
        Oxygen abundance & [O/H] & - & $1.3$ & $>0.0$ \\
        Nitrogen abundance & [N/H] & - & $-3.5$ & $<-1.2$ \\
        Sulfur abundance & [S/H] & - & $1.5$ &  $<1.5$ \\
        Volatile abundance & [(C+O)/H] &  - & $1.1$ & $>-0.2$ \\
    \enddata 
    \caption{List of parameters, priors, and results for atmospheric retrievals. Results are shown for the 6-component retrieval. The error bars on the retrieved medians correspond to the 68$\% / 1\sigma$ confidence interval.  Limits are given at 95\% confidence. A $*$ indicates the marginalized posterior for a parameter is unconstrained (i.e. the marginalized posterior spans the full range of the prior). In addition to these priors, we required that the atmospheric temperature stay below 4000 K at all pressure levels. The full corner plot is included in Appendix \ref{app:corner}. The derived abundance parameters are based on the gas abundances included in the retrieval, and do not include contributions from condensates or other species not included in the forward model. True atomic abundances are likely enhanced as a result of these effects.}
    \label{tab:priors}
\end{deluxetable*}

\subsection{Number of principal components}

We performed retrievals omitting 4, 6, and 8 principal components. Full corner plots for all retrievals are included in Appendix \ref{app:corner}. While in principle, the optimal number of omitted components is likely different between the $K$- and $L$-band epochs (and potentially between the two science fibers used in $L$), we omit the same number of components from each of the three time series for simplicity. Visual inspection of the residuals showed that omitting four or more components removed significant systematics for all orders of all three time series. 

The 4 component posterior prefers a cool $P-T$ profile with strong absorption features from CO and CH$_4$ and an absence of H$_2$O. This is potentially consistent with the carbon-rich chemistry described by \citet{giacobbe2021}, but is at odds with more recent results from \textit{JWST} \citet{xue2024}. The six and eight component cases, in contrast, prefer a H$_2$O-rich spectrum, with upper limits on all other species, consistent with \citet{xue2024}. The $P-T$ profile in these cases is also more consistent with expectations based on the equilibrium temperature of \hdb.

To clarify the 4 component case, we performed retrievals on the $K$ and $L$ band epochs separately for both 4 and 6 omitted components. A corner plot comparing the joint and individual retrievals is included in Appendix \ref{app:corner}. When we use only the $L$ band data with 4 omitted PCs, the resulting posterior is consistent with both the 6 component joint retrieval and the 6 component $L$ band retrieval. The 4 component joint retrieval is dominated by the $K$ band data, which prefers the carbon-rich composition. In contrast, retrievals on the $K$ band data omitting 6 PCs are consistent with a H$_2$O-dominated atmosphere.

These results demonstrate that the posterior from the 4 component $K$ band data is inconsistent with both the $L$ band data and with increasing the number of omitted PCs. This strongly suggests that the $K$ band data retains significant residuals after omitting 4 PCs that bias the retrieval results, but which are dropped from the time series when omitting 6 PCs. In Section \ref{sec:disc}, we assess the impact of expanding the prior range to allow thermal inversions, and again find that the $K$ band data with 4 dropped PCs is an outlier, returning a chemically implausible posterior. 

Due to the issues with the 4 component $K$ band time series, we take the 6 component joint retrieval as our fiducial case. These results demonstrate the importance of consistency checks in high-resolution retrievals. The carbon-rich composition from the 4 component $K$ band retrieval is close to the \texttt{easyCHEM} \citep{lei2024} chemical equilibrium predictions for $\rm C/O\sim2,\ [Fe/H]\sim-0.5$, and cannot be rejected as clearly non-physical. However, bona fide atmospheric detections should produce posteriors which are consistent across epochs, bandpasses, and minor changes in detrending procedure. This is not the case for the results of the 4 component $K$ band retrieval, but is satisfied by the other retrievals preferring a H$_2$O-rich composition. 

\subsection{Kinematics}

Both epochs covered a similar pre-eclipse phase range and a relatively small velocity baseline of $\sim40$ \kms, leading to a degeneracy between the $K_p$ and $\Delta v_{sys}$ parameters (seen in Figure \ref{fig:kpvsys}). The retrieved kinematics in Table \ref{tab:priors} show \dvsys\ consistent with the nominal ephemeris and a minor offset to negative \kp\ values, $\Delta$\kp $=-23\pm13$ \kms. This corresponds to a blueshift with respect to the assumed ephemeris of $\sim 12$ \kms\ at the start of the $K$ band observations, decreasing to $5$ \kms\ by the end.

A day-to-night wind should produce a net redshift for observations taken near secondary eclipse. Similarly, if the observed emission is dominated by hotter afternoon/evening longitudes, the planetary rotation should result in a slight redshift in the net emission spectrum. Previously, \citet{beltz2021} found that using a spectral model based on 3D GCMs with a $\sim5$\kms east-to-west wind resulted in a significant increase improvement in the detection strength of \hdb\ in CRIRES data, suggesting that we should expect to see a redshift as a result of planetary winds/rotation.

Comparing separate analyses of the $K$ and $L$ band data provides a possible explanation. Figure \ref{fig:kpvsys} shows that the $K$ band detection has a slight offset to negative \dvsys\ values, whereas the $L$ band is consistent with the nominal ephemeris. This can also be seen in the corner plots included in Appendix \ref{app:corner}. The $L$ band peak is also significantly wider along the \dvsys\ axis compared to the $K$ band, consistent with the lower spectral resolution in $L$. The offset is roughly consistent with the preferred value for $v_{\rm rot} = 10.7^{+1.5}_{-1.6}$ \kms\ from the retrieval. This suggest that the retrieval is matching the $K$ band \kp\ and \dvsys, with a wide broadening kernel to reduce the discrepancy between $K$ and $L$.

The origin of the offset between bandpasses is unclear. Ephemeris errors can produce offsets in the \kpvsys\ space, but the similar phase coverage in both bandpasses and the short time span between observations should lead to a consistent offset from any such errors. Errors in the wavelength calibration for one or both data sets could be responsible, but we expect the wavelength solutions from the KPIC DRP to be accurate to $\sim0.1$\kms. The $\sim10$\kms\ offset is comparable to the instrument resolution and should be obvious upon inspection, but for both $K$ and $L$ bands the wavelength solutions appear to provide a good fit to the data. Moreover, the offset is consistent between the individual orders for each data set, and between the two science fibers in the $L$ band data set. Additional observations in the $L$ band should eventually improve the wavelength solutions for that bandpass, which may shed light on this apparent discrepancy.

While the origin of this offset is unclear, its impact on the interpretation of our results is minimal. Separate retrievals on the $K$ and $L$ band data return consistent posteriors, aside from \kp\ and \dvsys. Joint retrievals including a free parameter for the velocity offset between bands returned a constrained value for the offset of $11\pm2$ \kms, but were consistent with the retrievals excluding an offset for the all of the $P-T$ and abundance parameters.

As discussed in \citet{finnerty2025a, finnerty2025b}, our LSF model assumes a fixed factor of 1.14 in converting between the spatial and spectral instrument response profiles \citep{Finnerty2022}. This value is expected to be achromatic, but this has not been confirmed in the $L$ band specifically. Our measurements find the NIRSPEC LSF in the $L$ band is significantly broader than the $K$ band, consistent with expectations from the instrument design, resulting in the reduced spectral resolution. Given the systematic uncertainties involved in the LSF fitting, we are wary of physical interpretation of the rotational broadening parameter, even in the absence of the \kpvsys\ discrepancy between bandpasses. Constraining wind speeds from emission observations will likely require significantly broader phase coverage, in order to separate the effects of instrumental and ephemeris uncertainties from non-Keplerian motion associated with circulation patterns.

\subsection{Pressure-temperature profile}\label{ssec:thermal}

\begin{figure*}
    \centering
    \includegraphics[width=0.9\linewidth]{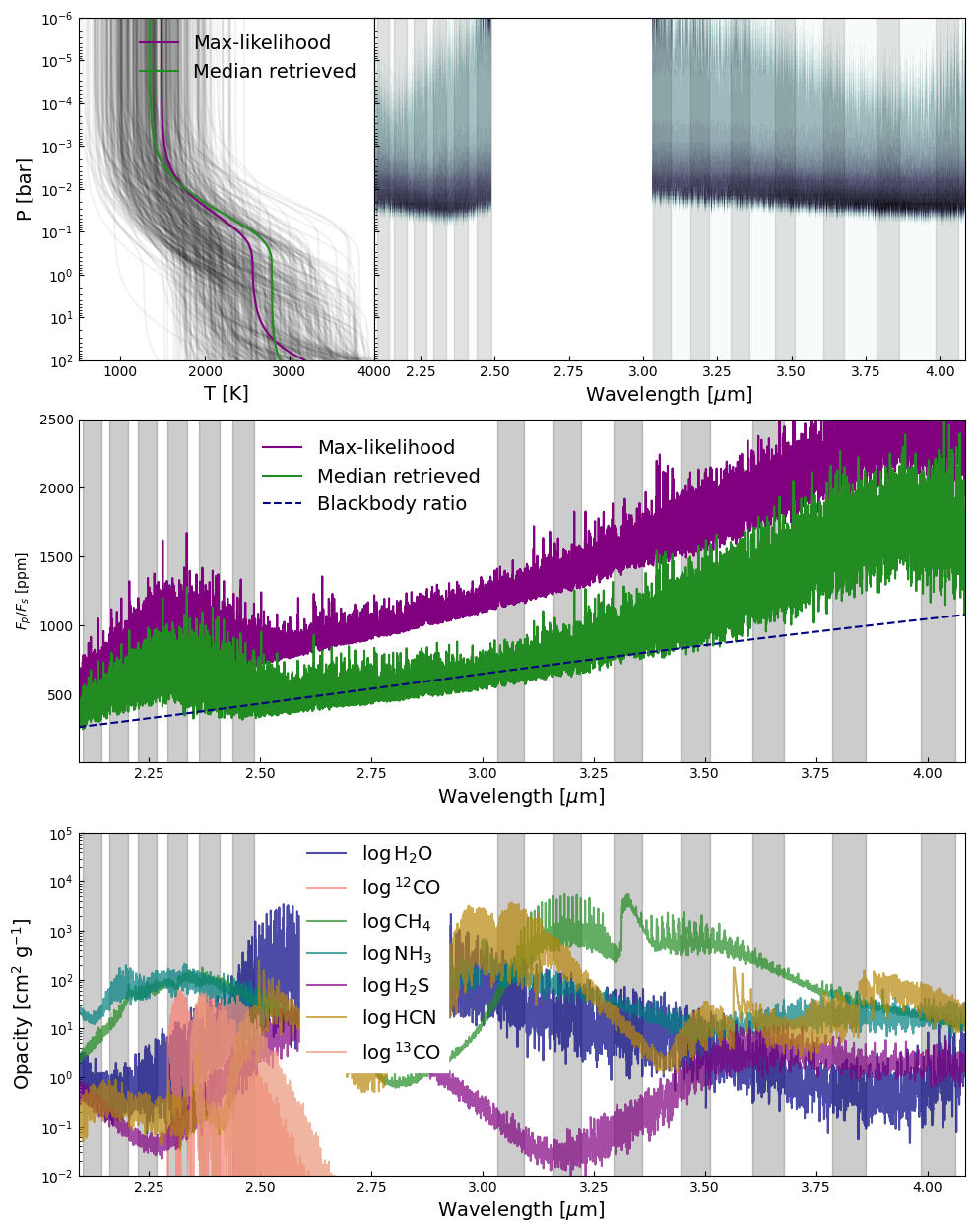}
    \caption{Top left: retrieved median and max-likelihood $P-T$ profiles, with 500 $P-T$ profiles drawn from the posterior in black. Top right: Maximum-likelihood emission contribution function, with observed orders shaded. Center: $F_p/F_s$ for the maximum likelihood and median retrieved planet models, compared with the expected $F_p/F_s$ assuming the planet and star are blackbodies at their respective equilibrium and effective temperatures. Observed orders are shaded. Bottom: Opacities of the species included in the retrieval over the observed bandpass.}
    \label{fig:specplot}
\end{figure*}

Figure \ref{fig:specplot} shows the retrieved $P-T$ profile, max-likelihood emission contribution, retrieved $F_p/F_s$, and smoothed opacities of the species included. While the median and maximum-likelihood values for $\rm T_{equ}$ are somewhat higher than the calculated $1450\rm\ K$ equilibrium temperature of \hdb, Figure \ref{fig:specplot} shows the resulting $F_p/F_s$ for the median and maximum-likelihood models is roughly consistent with the expected values assuming both star and planet are blackbodies.

Compared with KPIC $K$ band HRCCS studies \citep{finnerty2023, finnerty2024, finnerty2025b}, our results are much more consistent with the expected temperature of \hdb\ and a scale factor of 1. The wider spectral coverage from combining $K$ and $L$ band data appears to result in significantly improved constraints on the $P-T$ profile, avoiding the degeneracy between the $P-T$ contrast and scaling parameter discussed in those works. This is consistent with results from IGRINS \citep[e.g.]{line2021}, which covers the full $H$ and $K$ bands simultaneously and provides tight constraints on the $P-T$ profile in retrievals. 

\subsection{Molecular abundances}\label{ssec:chem}

\kpvsys\ plots for individual molecules are shown in Figure \ref{fig:kpvsysmols}. For each molecule, we created a template using the maximum-likelihood parameters, but set the mass fraction of the molecule in question to $10^{-1.5}$ and the mass fraction of other species to $10^{-15}$. We then calculate the log-likelihood over a grid of \kp\ and \dvsys\ values to make the \kpvsys\ map. We convert this map to signal-to-noise by first median-subtracting each row, to remove the systematic effect with \kp\ described in \citet{finnerty2025b}, then divide the entire map by the standard deviation of the negative \kp\ region. See \citet{finnerty2024, finnerty2025b} for further discussion of noise estimation in \kpvsys\ maps.

Consistent with the retrieved posterior, only the H$_2$O template returns a detection of \hdb. While the H$_2$S template shows an $\rm SNR\sim3$ feature at the expected \kp and \dvsys\ values, features of similar strength are present throughout the \kpvsys\ space, and this does not constitute a detection of H$_2$S. 

The retrieval returns a lower limit on the H$_2$O mass fraction ($>10^{-2.2}$ at 95\% confidence), and upper limits on all other species. Table \ref{tab:priors} shows the $95\%$ confidence limits on the mass-mixing ratio for each species from the 6 PC joint retrieval. The lower limit on H$_2$O places a lower limit on the oxygen abundance $\rm [O/H] > 0$ at 95\% confidence, assuming the \citet{asplund2021} solar elemental abundances.

If significant quantities of CH$_4$ were present in the atmosphere of \hdb, it should result in significant absorption features in the three observed $L$-band orders from 3.2--3.5 $\mu$m. The apparent absence of these features results in a strong upper limit on the CH$_4$ mass fraction, $<10^{-3.9}$ at 95\% confidence. The limit on the CO mass fraction is similar, $<10^{-3.7}$ at 95\% confidence. These abundances imply a significantly depleted gas-phase carbon abundance compared to the solar value, $[\rm C/H] < -0.9$ at 95\% confidence.

As shown in Figure \ref{fig:specplot}, the observed orders include strong opacity features from NH$_3$ at the blue end of the $K$ band and HCN at either end of the $L$ band. Despite these strong features, the retrieval prefers mass fractions $\lesssim10^{-4.5}$ for both species. If these species are the dominant gas-phase reservoirs of nitrogen, this would suggest significant nitrogen depletion relative to solar. However, the abundances listed in Table \ref{tab:priors} are based only on the detected gas species, and do not account for nitrogen in condensed species and/or in gas species which were not included in the retrieval.

Compared to the other species considered, the constraint on the H$_2$S mass fraction is relatively poor, $\rm H_2S_{MMR} < 10^{-2.1}$ at 95\% confidence. This weak limit permits significant sulfur enrichment compared to the solar value, $\rm [S/H] < 1.5$ at 95\% confidence. However, H$_2$S is not significantly detected in cross-correlation, nor is it contained in the retrievals. Figure \ref{fig:specplot} shows that, in contrast to the other species included in the retrieval, at no point in the observed bandpasses is H$_2$S opacity dominant. While the H$_2$S opacity is slightly greater than H$_2$O at the longest wavelengths, the much higher abundance of H$_2$O will still result in a H$_2$O-dominated spectrum in this region, making comparatively tiny H$_2$S features difficult to distinguish. Constraints on the H$_2$S abundance in \hdb\ will likely require observations in bandpasses with significantly stronger H$_2$S features.

\section{Discussion}\label{sec:disc}

\subsection{Ruling out thermal inversions}

To date, thermal inversions have not been observed in hot Jupiters with equilibrium temperatures $<2000$ K, and several hot Jupiters with $1600 < \rm T_{equ} < 2000$ have been confirmed to lack thermal inversions \citep[e.g.][]{line2021, finnerty2025a}. Recent modeling work has found that thermal inversions only form at temperatures $\gtrsim1800$ K  in the presence of significant TiO/VO \citep{roth2024}, with significantly higher temperatures required in the absence of these species \citep{lothringer2018, baxter2020}. These findings suggest \hdb\ should not have a significant thermal inversion. Observationally, \citet{line2016} used a combination of \textit{HST} and \textit{Spitzer} observations to rule out thermal inversions in \hdb, following high-resolution observations reported \citet{schwarz2015} which excluded a thermal inversion based on a non-detection of CO emission features. These results motivated our choice to exclude thermal inversions from the prior range of our fiducial retrievals.

However, earlier studies claimed detection of a thermal inversion in \hdb\ based on \textit{Spitzer} secondary eclipse observations \citep{burrows2007, knutson2008}. While atmospheric models including TiO/VO could produce an inversion in \hdb, these models failed to agree with the \textit{Spitzer} photometry \citep{showman2009}. Subsequent analysis of \textit{Spitzer} data suggested that the points driving the thermal inversion were a result of systematics \citep{diamondlowe2014, line2016}. 

For completeness, we ran a set of retrievals with an expanded prior $-3 < \log\gamma < 3$, which permits thermal inversions. The posterior for the six component joint retrieval was not significantly changed compared with the results presented in Table \ref{tab:priors}, indicating the KPIC high-resolution data strongly reject a thermal inversion. The $K$ band data with 4 omitted PCs prefers a strong inversion with CH$_4$ emission features, in tension with chemical equilibrium predictions and previous results, further demonstrating that this time series is strongly impacted by non-planetary residuals. 

\subsection{Atmospheric composition}

Previous results in the literature have typically provided molecular abundances as Volume Mixing Ratios (VMRs), rather than the mass-mixing ratios used in our retrieval pipeline. Assuming a mean molecular weight of 2.3 g/mol, the retrieved H$_2$O abundance corresponds to a $\rm \log H_2O_{VMR} >-3.1$. The 95\% upper limits correspond to $\rm \log CO_{VMR} < -4.8$, $\rm \log CH_{4,VMR} < -4.8$, $\rm \log NH_{3,VMR} < -5.8$, $\rm \log H_2S_{VMR} < -3.3$, and $\rm \log HCN_{VMR} < -5.6$. Our retrieved 99.7\%/$3\sigma$ limits on the VMRs are $\rm \log H_2O_{VMR} > -3.9$, $\rm \log CO_{VMR} < -2.8$, $\rm \log CH_{4,VMR} < -3.8$, $\rm \log NH_{3,VMR} < -4.7$, $\rm \log H_2S_{VMR} < -2.4$, and $\rm \log HCN_{VMR} < -3.7$.  Our broad wavelength coverage allows us to place these  limits on the chemical composition of \hdb\ with only two half nights of ground-based observations. 

The 95\% upper limit on the CO abundance is difficult to reproduce with chemical equilibrium models. Comparing the retrieved 95\% limits to predictions from \texttt{easyCHEM} \citep{lei2024} requires $\rm [Fe/H] < -1.3$ and $\rm C/O < 2\times10^{-3}$. This is in tension with the other recent observations of \hdb\ discussed below. However, the 3$\sigma$/99.7\% limit on the CO abundance is approximately 2 dex greater, and a CO abundance between the 95\% and 99.7\% limit would be consistent with $\rm [Fe/H] \sim0.5$ and $\rm C/O \sim0.1$ and compatible with the results of other studies.

Figure \ref{fig:specplot} shows that even at the 2.3$\mu$m bandhead, the CO opacity is within an order of magnitude of the H$_2$O opacity. For an atmosphere with $\rm C/O\lesssim0.1$, the spectrum could be dominated by H$_2$O features at the CO bandhead. If the comparatively weak CO features are close to the noise limit of the data, including significant quantities of CO in the forward model could lead to a marginal improvement in the fit which is insufficient to overcome the model variance penalty of the \citet{brogi2019} log-likelihood function. This could bias the retrieval to stricter limits on the CO abundance. Alternatively, our observations covered a velocity shift only $\sim4.5\times$ the spectral resolution of the KPIC/NIRSPEC system, and the regular structure of the $\rm 2.3\mu m$ CO bandhead may make CO features particularly vulnerable to self-subtraction during PCA. Data with a wider phase coverage and increased sensitivity are needed to clarify the CO abundance in \hdb. Future analyses incorporating existing CRIRES $K$ band observations may enable this without new observations. 

\subsection{Comparison to previous results}

Recently, \citet{xue2024} detected strong H$_2$O and CO$_2$ features in the transmission spectrum of \hdb\ observed with \textit{JWST}, possible weak CO and H$_2$S features, and placed 3$\sigma$ upper limits of $\rm \log CH_{4,VMR} < -5.6$, $\rm \log NH_{3,VMR} < -4.2$, and $\rm \log HCN_{VMR} < -5.1$. The bulk composition was best fit by a low C/O ratio ($\sim0.1$) and moderately metal-enriched ($\sim3\times$ solar) atmosphere \citep{xue2024}. Similar results were obtained by \citet{bachman2025}, who combined the \textit{JWST} data with re-reduced \textit{HST} transmission observations, finding a roughly solar metallicity ($\rm [M/H] = 0.1\pm0.4$) and very low $\rm C/O = 0.05^{+0.08}_{-0.03}$. These results are broadly consistent with the retrieval analysis presented here, given the above discussion of the retrieved CO abundance. 
  
At high spectral resolution, \citet{giacobbe2021} reported detections of CO, HCN, CH$_4$, NH$_3$, and C$_2$H$_2$ in the atmosphere of \hdb from GIANO-B high-resolution transmission observations, which cover $0.95-2.45\rm\ \mu m$. \citet{hawker2018} reported detections of H$_2$O, CO, and HCN from a combination of $K$- and $L$-band emission data, and \citet{snellen2010} reported a detection of CO in $K$-band transmission data. However, both \citet{cheverall2023} and \citet{blain2024hd209} recently detected only H$_2$O in the atmosphere of \hdb\ at high spectral resolution, consistent with our analysis and with the results presented in \citet{xue2024}.

As discussed above, the non-detection of CO in our analysis is likely due to some combination of the limited phase coverage of our observations and relatively weak CO features in the low C/O atmosphere of \hdb, which is potentially compounded by the model variance penalty in the log-likelihood function. The spectral resolution of the observations presented by \citet{snellen2010} was approximately $3\times$ greater than our observations, with much better ability to detect CO as a result.

In emission, \citet{schwarz2015} initially reported a non-detection of CO in \hdb\ from CRIRES observations. Subsequent reanalyses of these data presented in \citet{brogi2019} and \citet{gandhi2019} both reported a CO detection in these data. Both retrieval analyses preferred a CO-rich atmosphere, with \citet{brogi2019} placing a subsolar upper limit on the H$_2$O abundance and \citet{gandhi2019} constraining H$_2$O to a very low value. However, \citet{hawker2018} reported a detection of H$_2$O absorption features in these same data, and previous low-resolution observations of \hdb\ preferred a solar H$_2$O abundance \citep{line2016}. 

\citet{brogi2019} also analyzed CRIRES observations of HD~189733~b, finding a CO-rich composition similar to their results for \hdb. Subsequent ground-based high-resolution and \textit{JWST} observations of HD~189733~b have preferred a low C/O ratio and moderately super-solar metallicity for HD~189733~b \citep{finnerty2024, fu2024}. This similar discrepancies between \citet{brogi2019} and subsequent analyses for HD~189733~b and \hdb\ suggests a systematic underestimation of the H$_2$O abundance in the \citet{brogi2019} retrievals. 

\citet{brogi2019} ran separate retrievals with H$_2$O cross-sections from HITEMP \citep{hitemp2010} and from \citet{freedman2014} based on the \citet{partridge1997} linelist. Those authors note significant differences in the final H$_2$O spectrum between the two linelists, which biased retrieved abundances when one linelist was used to generate synthetic data and the other was used for the retrieval. However, neither linelist gave a constrained H$_2$O abundance in their retrievals on CRIRES observations.  In the case of $K$ band CRIRES transmission observations of HD~189733~b, \citet{flowers2019} found H$_2$O was detected using HITEMP opacities, but not with opacities from \citet{lupu2014}. 

\citet{finnerty2024} measured the H$_2$O abundance in HD~189733~b with opacities generated using \texttt{exocross} \citep{exocross2018} based on the \citet{hitemp2010} linelist and the \citet{polyansky2018} partition function, obtaining a value consistent with subsequently published \textit{JWST} observations \citep{fu2024}. For our analysis of \hdb\, we used opacities based on the  \citet{polyansky2018} linelist/partition function, again obtaining results consistent with \textit{JWST} \citep{xue2024}. These results suggest that linelist inaccuracies may have significantly impacted H$_2$O abundances in retrieval studies prior to 2020, but that more recent linelists are more accurate. 

Alternatively, the CRIRES observations used in \citet{brogi2019} covered only $\sim40$ nm around the $2.3\mu\rm m$ CO bandhead. Over this narrow range, shielding from strong CO absorption may effectively suppress H$_2$O features, resulting in retrievals preferring the presence of CO alone. In contrast, the broad bandpass data presented here offer good sensitivity to H$_2$O features throughout both the $K$ and $L$ bands, but the small phase coverage may make CO features particularly vulnerable to subtraction during PCA. The self-subtraction scenario is consistent with the relatively weak upper limit for CO compared with other species. 

The CH$_4$ and NH$_3$, and HCN detections reported in \citet{giacobbe2021} and HCN detection from \citet{hawker2018} may be false positives as a result of detrending procedures. In the case of HCN, \citet{savel2025} recently demonstrated that HCN may be particularly vulnerable to false positive detections in high-resolution spectroscopy. As discussed in \citet{cheverall2023} and \citet{blain2024hd209}, detrending or order selection procedures based on optimizing a planet signal, such as those used in \citet{hawker2018} and \citet{giacobbe2021}, significantly increase the risk of false positive molecular detections. The case of our $K$ band observations with 4 PCs omitted is an instructive example of the possible failure modes. The resulting posterior prefers a  high C/O ratio and low metallicity, which is not obviously non-physical, but which is quickly rejected by comparing the results of minor changes in detrending procedure and/or comparing the results from observations in multiple bandpasses. This may not have been noticed if only a single set of PCs was used and the included orders were chosen to maximize detection strength, or if only $K$ band data were available. High-resolution retrieval analyses should attempt to demonstrate robustness to minor changes in detrending procedures and/or consistency across multiple epochs/bandpasses, particularly when novel or unexpected planetary properties are suspected. 

\subsection{Formation history}

The retrieved molecular abundances listed in Table \ref{tab:priors} suggest a gas-phase metallicity (in this case [(C+O)/H], rather than [Fe/H]) at of least approximately solar, accompanied by significant carbon depletion resulting in a very low C/O ratio. The values reported in Table \ref{tab:priors} include only the molecular species included in the retrieval, and the true bulk atmospheric abundances will be different due to contributions from excluded species and condensates. The impact of these exclusions can be assessed by comparing the retrieved abundances to chemical equilibrium models including a wider range of molecular species and condensates.

As previously discussed, matching the retrieved 95\% upper limits to equilibrium chemistry predictions from \texttt{easyCHEM} \citep{lei2024} requires very low atmospheric metallicity and C/O ratio due to the low CO abundance. Such a composition is in tension with results from \textit{JWST} \citep{xue2024}. The 99.7\% limit on CO is compatible with significantly greater metallicities, and is consistent with the $\rm C/O\sim0.1,\ [Fe/H]\sim+0.5$ results from \citet{xue2024}. The equilibrium CO abundance is below the 99.7\% upper limit for $\rm [Fe/H] < 1.0$. Equilibrium abundances for the other species included in the retrieval are well below the 95\% upper limits even at $10\times$ solar metallicity. Disequilibrium chemistry involving these species may nevertheless be occurring in the atmosphere of \hdb, if the resulting abundances are still below our detection limits. 

The poor metallicity constraints retrieved from the data presented here limit inferences about the formation history of \hdb\ form these data alone. In general, C/O and metallicity are expected to be inversely correlated in core-accretion scenarios \citep[e.g.][]{espinoza2017, madhusudhan2017, cridland2019}. The low C/O ratio is consistent with formation scenarios that involve late accretion of refractory solids and/or ices after most of the gas has been accreted \citep{oberg2011, khorshid2022}, which would also produce a stellar to slightly super-stellar metallicity, which is consistent with our retrievals and the metallicity obtained by \citet{xue2024}. 

\section{Summary and Conclusions}\label{sec:conc}

We performed a joint retrieval analysis on two epochs of KPIC observations of \hdb, covering pre-eclipse phases in both $K$ and $L$ bands. This analysis prefers a low gas-phase C/O ratio ($<10^{-3}$ at 95\% confidence), and volatile metallicity $\rm [(C+O)/H] > -0.2$ at 95\% confidence. These results are consistent with low-resolution transmission spectroscopy from \textit{JWST} \citep{xue2024}. The wide wavelength coverage of our observations enable us to place strict upper limits on the abundances of CH$_4$, NH$_3$, H$_2$S, and HCN with only two half nights of observations. These limits are consistent with recent results from \textit{JWST} and CRIRES+ in transmission, but in tension with previous ground-based high-resolution detections of these species, which took more aggressive approaches to detrending. The retrieved composition of \hdb\ is broadly consistent with chemical equilibrium, and the low C/O ratio is consistent with significant accretion of oxygen-rich planetesimals late in the planet formation process.

\begin{acknowledgments}
L. F. is a member of UAW local 4811. L.F. acknowledges the support of the W.M. Keck Foundation, which also supports development of the KPIC facility data reduction pipeline. The contributed Hoffman2 computing node used for this work was supported by the Heising-Simons Foundation grant \#2020-1821. Funding for KPIC has been provided by the California Institute of Technology, the Jet Propulsion Laboratory, the Heising-Simons Foundation (grants \#2015-129, \#2017-318, \#2019-1312, \#2023-4597, \#2023-4598), the Simons Foundation (through the Caltech Center for Comparative Planetary Evolution), and the NSF under grant AST-1611623. D.E. acknowledges support from the NASA Future Investigators in NASA Earth and Space Science and Technology (FINESST) fellowship under award \#80NSSC19K1423, as well as support from the Keck Visiting Scholars Program (KVSP) to install the Phase II upgrades for KPIC. J.X. acknowledges support from the NASA Future Investigators in NASA Earth and Space Science and Technology (FINESST) award \#80NSSC23K1434.

This work used computational and storage services associated with the Hoffman2 Shared Cluster provided by UCLA Institute for Digital Research and Education’s Research Technology Group. L.F. thanks Briley Lewis for her helpful guide to using Hoffman2, and Paul Molli\`ere for his assistance in adding additional opacities to petitRADTRANS. 

The data presented herein were obtained at the W. M. Keck Observatory, which is operated as a scientific partnership among the California Institute of Technology, the University of California and the National Aeronautics and Space Administration. The Observatory was made possible by the generous financial support of the W. M. Keck Foundation. W. M. Keck Observatory access was supported by Northwestern University and the Center for Interdisciplinary Exploration and Research in Astrophysics (CIERA). The authors wish to recognize and acknowledge the very significant cultural role and reverence that the summit of Mauna Kea has always had within the indigenous Hawaiian community.  We are most fortunate to have the opportunity to conduct observations from this mountain. 

This research has made use of the NASA Exoplanet Archive \citep{10.26133/nea12} and the Exoplanet Follow-up Observing Program \citep{10.26134/ExoFOP5}, which are operated by the California Institute of Technology, under contract with the National Aeronautics and Space Administration under the Exoplanet Exploration Program. The research shown here acknowledges use of the Hypatia Catalog Database, an online compilation of stellar abundance data as described in Hinkel et al. (2014, AJ, 148, 54), which was supported by NASA's Nexus for Exoplanet System Science (NExSS) research coordination network and the Vanderbilt Initiative in Data-Intensive Astrophysics (VIDA).

\end{acknowledgments}

%

\vspace{5mm}
\facilities{Keck:II(NIRSPEC/KPIC)}


\software{astropy \citep{astropy:2013, astropy:2018},  
          \texttt{corner} \citep{corner},
          \petit\ \citep{prt:2019, prt:2020}}


\appendix
\section{Corner Plots}\label{app:corner}

\begin{figure}
    \centering
    \includegraphics[width=1.0\linewidth]{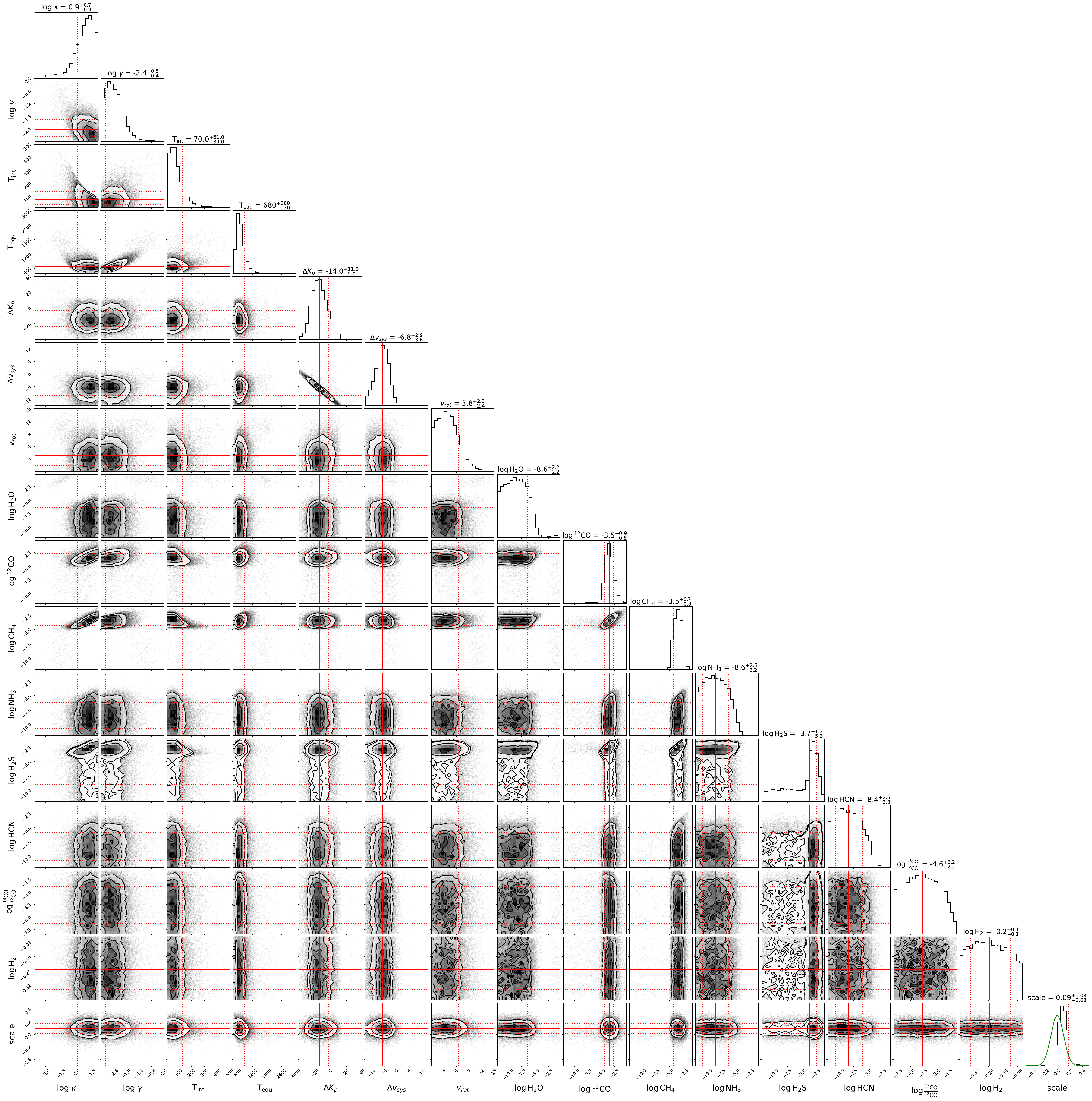}
    \caption{Full corner plot for the 4-component joint retrieval.}
    \label{fig:corner4}
\end{figure}

\begin{figure}
    \centering
    \includegraphics[width=1.0\linewidth]{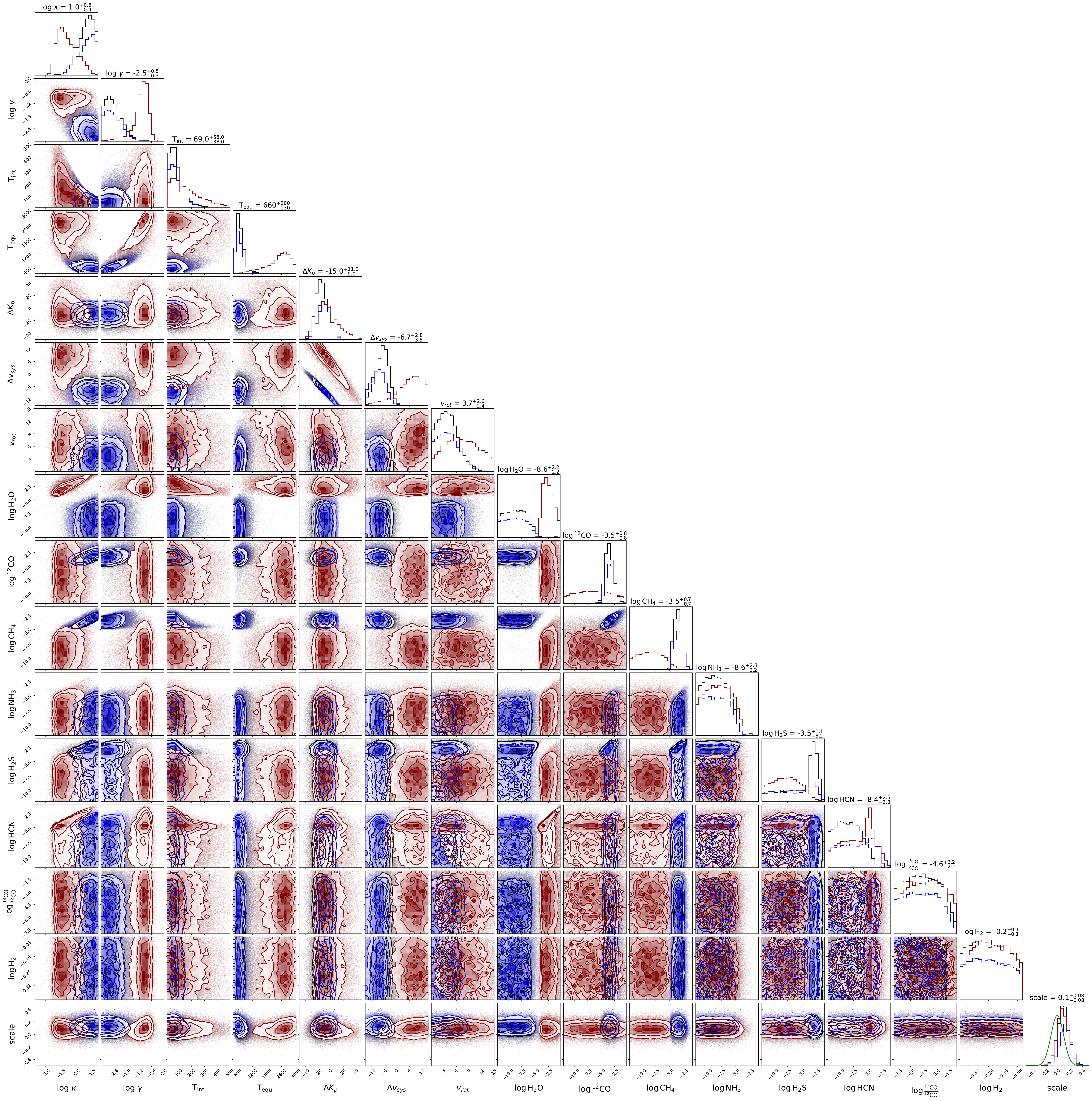}
    \caption{Full corner plot comparing the 4-component joint retrieval with the results from each band separately. While the $L$ band is consistent with the results from the 6-component joint retrieval, the $K$ band prefers a carbon-rich composition, and dominates the joint fit. Comparison with the $L$ band and 6 and 8 component cases indicates that this mode is a result of residuals in the $K$ band time series when omitting 4 PCs..}
    \label{fig:corner4comp}
\end{figure}

\begin{figure}
    \centering
    \includegraphics[width=1.0\linewidth]{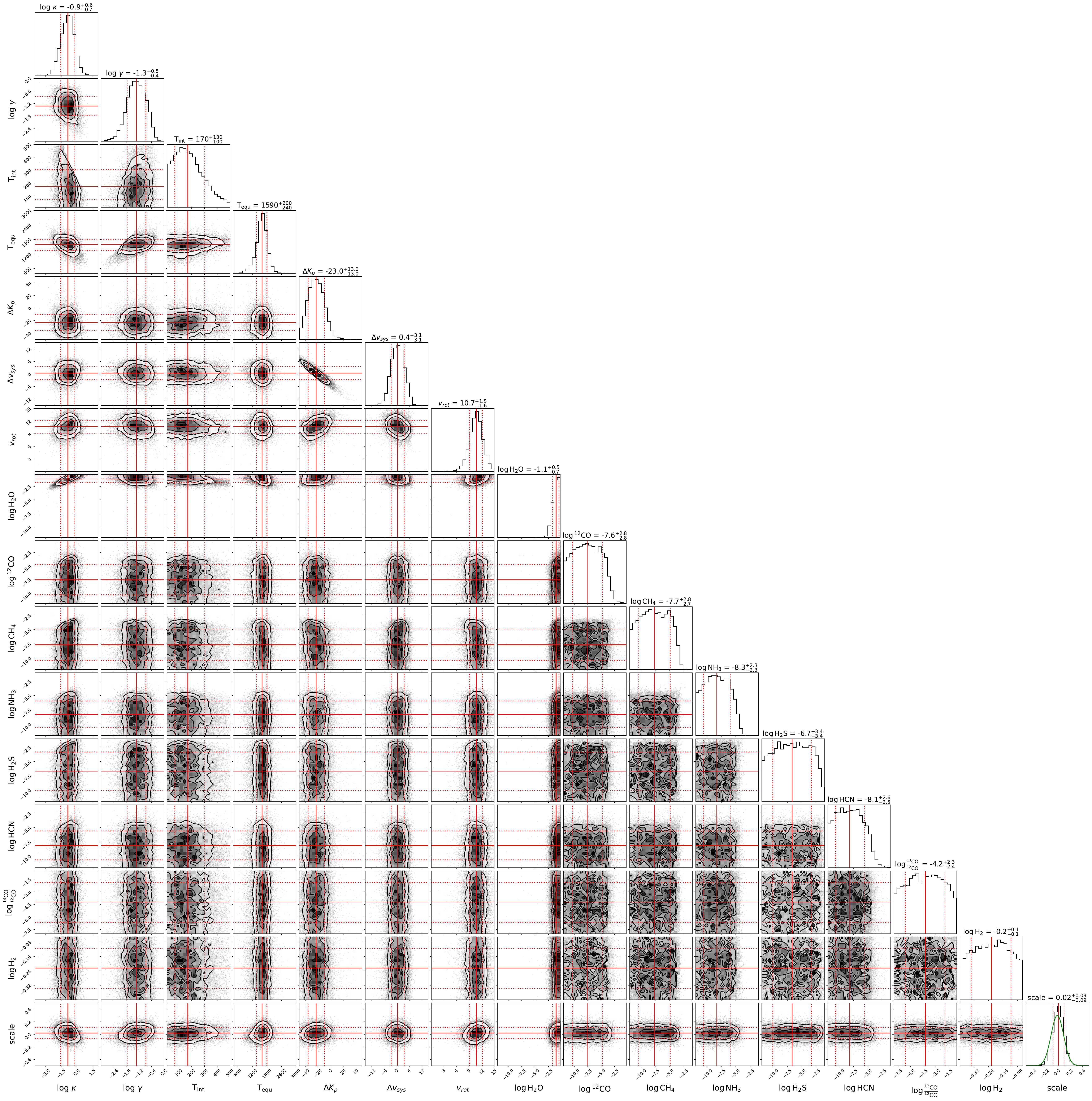}
    \caption{Full corner plot for the 6-component joint retrieval.}
    \label{fig:corner6}
\end{figure}

\begin{figure}
    \centering
    \includegraphics[width=1.0\linewidth]{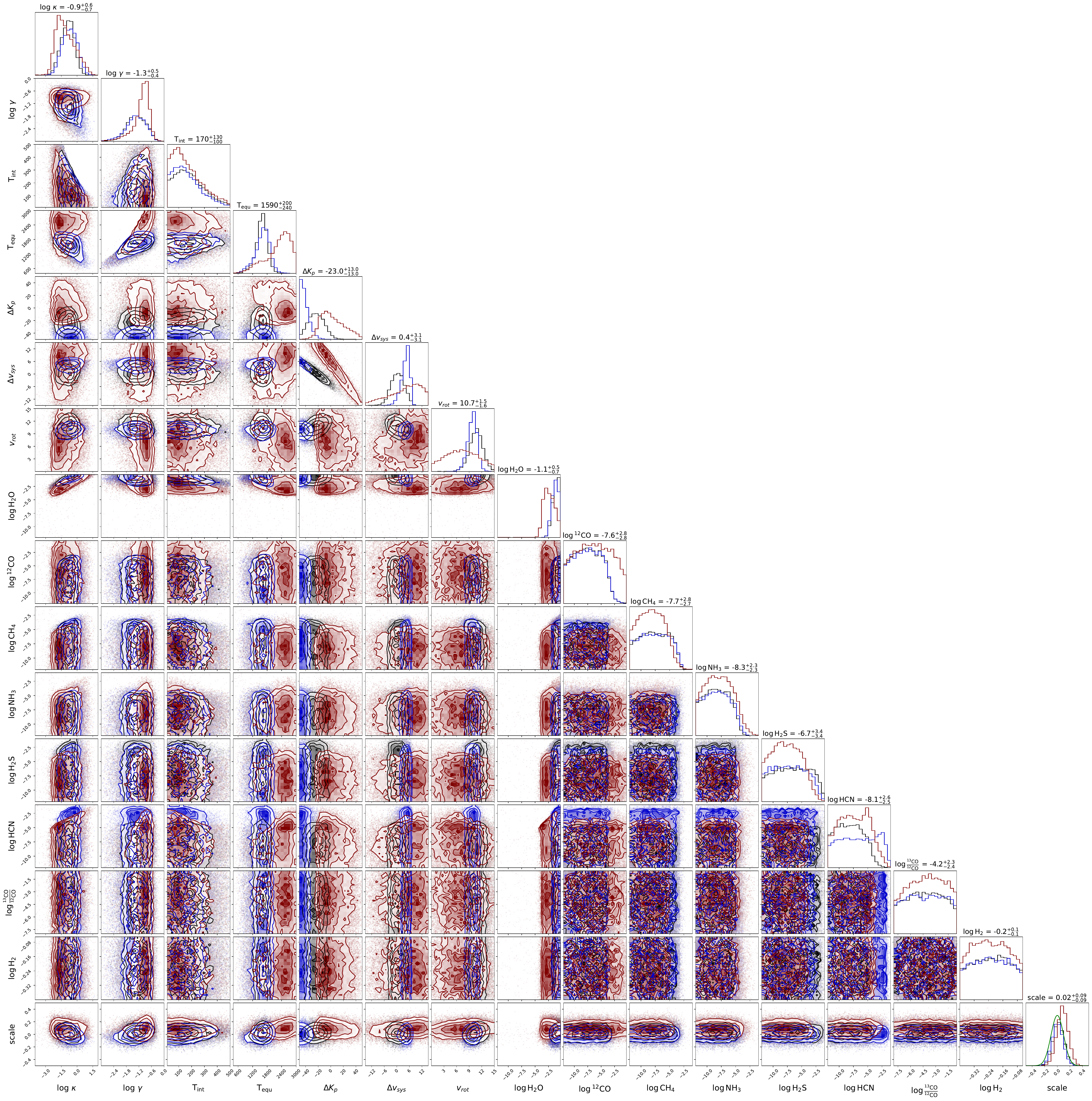}
    \caption{Full corner plot comparing the 6-component joint retrieval with the results from each band separately. The posteriors are generally consistent, except for a slight offset in the \kp\ and \dvsys\ parameters. This could be explained by errors in the $L$ band wavelength calibration, and may be responsible for the relatively large value of $v_{\rm rot}$. }
    \label{fig:corner6comp}
\end{figure}

\begin{figure}
    \centering
    \includegraphics[width=1.0\linewidth]{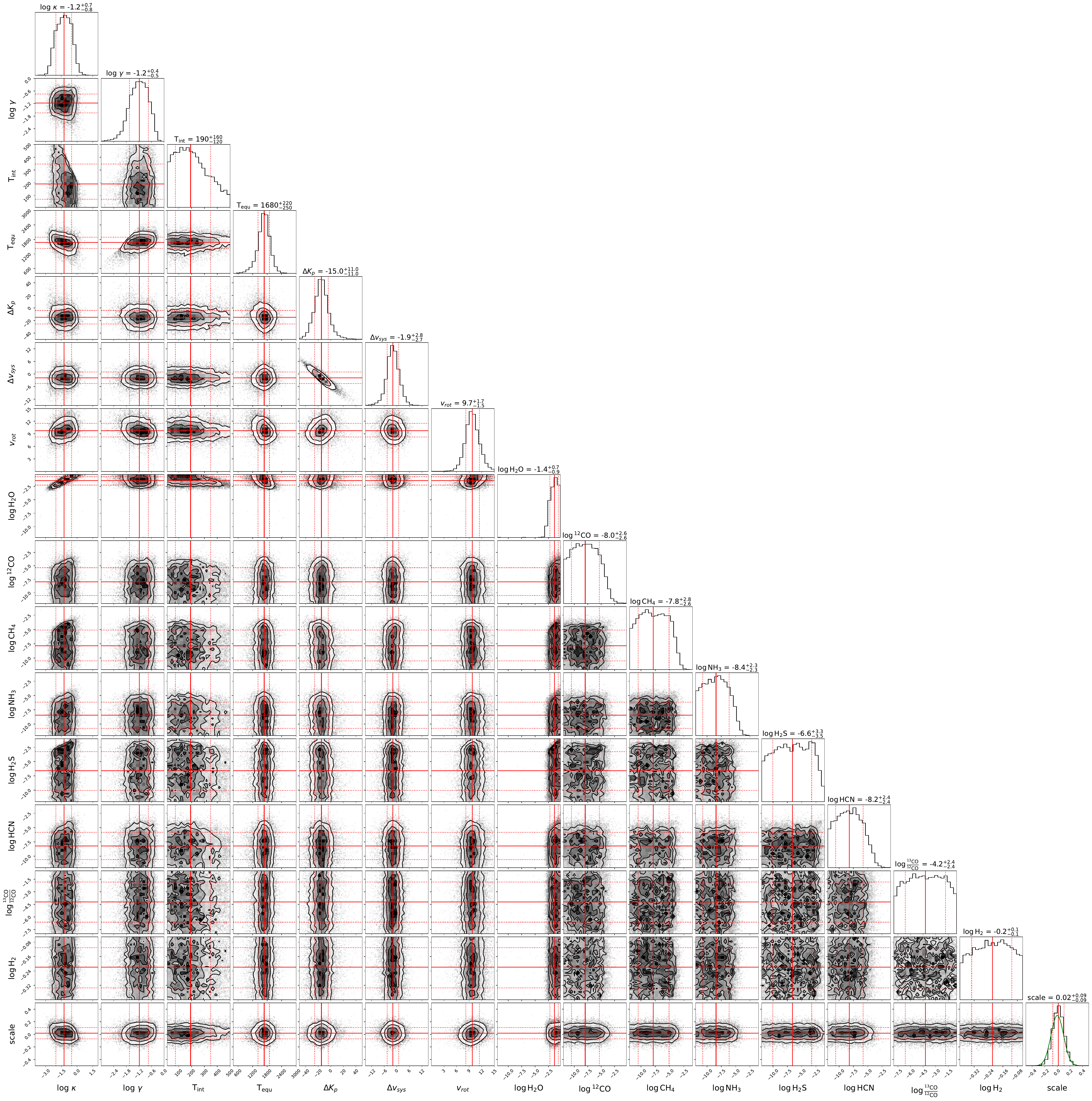}
    \caption{Full corner plot for the 8-component joint retrieval.}
    \label{fig:corner8}
\end{figure}

Figure \ref{fig:corner4} presents the full corner plot from the joint retrieval with 4 principal components omitted. Figure \ref{fig:corner4comp} compares the 4 PC joint retrieval with separate retrievals for the $K$ and $L$ band data omitting 4 PCs. Figure \ref{fig:corner6} presents the full corner plot from the 6 PC joint retrieval, which is the fiducial retrieval presented in Table \ref{tab:priors}. Figure \ref{fig:corner6comp} is analogous to Figure \ref{fig:corner4comp}, but with 6 PCs omitted. Finally, Figure \ref{fig:corner8} presents the full corner plot from the 8 PC joint retrieval. 

\clearpage
\bibliography{exoplanetbib}{}

\begin{thebibliography}{}
\expandafter\ifx\csname natexlab\endcsname\relax\def\natexlab#1{#1}\fi
\providecommand{\url}[1]{\href{#1}{#1}}
\providecommand{\dodoi}[1]{doi:~\href{http://doi.org/#1}{\nolinkurl{#1}}}
\providecommand{\doeprint}[1]{\href{http://ascl.net/#1}{\nolinkurl{http://ascl.net/#1}}}
\providecommand{\doarXiv}[1]{\href{https://arxiv.org/abs/#1}{\nolinkurl{https://arxiv.org/abs/#1}}}

\bibitem[{{Asplund} {et~al.}(2021){Asplund}, {Amarsi}, \& {Grevesse}}]{asplund2021}
{Asplund}, M., {Amarsi}, A.~M., \& {Grevesse}, N. 2021, \aap, 653, A141, \dodoi{10.1051/0004-6361/202140445}

\bibitem[{{Astropy Collaboration} {et~al.}(2013){Astropy Collaboration}, {Robitaille}, {Tollerud}, {Greenfield}, {Droettboom}, {Bray}, {Aldcroft}, {Davis}, {Ginsburg}, {Price-Whelan}, {Kerzendorf}, {Conley}, {Crighton}, {Barbary}, {Muna}, {Ferguson}, {Grollier}, {Parikh}, {Nair}, {Unther}, {Deil}, {Woillez}, {Conseil}, {Kramer}, {Turner}, {Singer}, {Fox}, {Weaver}, {Zabalza}, {Edwards}, {Azalee Bostroem}, {Burke}, {Casey}, {Crawford}, {Dencheva}, {Ely}, {Jenness}, {Labrie}, {Lim}, {Pierfederici}, {Pontzen}, {Ptak}, {Refsdal}, {Servillat}, \& {Streicher}}]{astropy:2013}
{Astropy Collaboration}, {Robitaille}, T.~P., {Tollerud}, E.~J., {et~al.} 2013, \aap, 558, A33, \dodoi{10.1051/0004-6361/201322068}

\bibitem[{{Astropy Collaboration} {et~al.}(2018){Astropy Collaboration}, {Price-Whelan}, {Sip{\H{o}}cz}, {G{\"u}nther}, {Lim}, {Crawford}, {Conseil}, {Shupe}, {Craig}, {Dencheva}, {Ginsburg}, {Vand erPlas}, {Bradley}, {P{\'e}rez-Su{\'a}rez}, {de Val-Borro}, {Aldcroft}, {Cruz}, {Robitaille}, {Tollerud}, {Ardelean}, {Babej}, {Bach}, {Bachetti}, {Bakanov}, {Bamford}, {Barentsen}, {Barmby}, {Baumbach}, {Berry}, {Biscani}, {Boquien}, {Bostroem}, {Bouma}, {Brammer}, {Bray}, {Breytenbach}, {Buddelmeijer}, {Burke}, {Calderone}, {Cano Rodr{\'\i}guez}, {Cara}, {Cardoso}, {Cheedella}, {Copin}, {Corrales}, {Crichton}, {D'Avella}, {Deil}, {Depagne}, {Dietrich}, {Donath}, {Droettboom}, {Earl}, {Erben}, {Fabbro}, {Ferreira}, {Finethy}, {Fox}, {Garrison}, {Gibbons}, {Goldstein}, {Gommers}, {Greco}, {Greenfield}, {Groener}, {Grollier}, {Hagen}, {Hirst}, {Homeier}, {Horton}, {Hosseinzadeh}, {Hu}, {Hunkeler}, {Ivezi{\'c}}, {Jain}, {Jenness}, {Kanarek}, {Kendrew}, {Kern}, {Kerzendorf}, {Khvalko}, {King}, {Kirkby}, {Kulkarni},
  {Kumar}, {Lee}, {Lenz}, {Littlefair}, {Ma}, {Macleod}, {Mastropietro}, {McCully}, {Montagnac}, {Morris}, {Mueller}, {Mumford}, {Muna}, {Murphy}, {Nelson}, {Nguyen}, {Ninan}, {N{\"o}the}, {Ogaz}, {Oh}, {Parejko}, {Parley}, {Pascual}, {Patil}, {Patil}, {Plunkett}, {Prochaska}, {Rastogi}, {Reddy Janga}, {Sabater}, {Sakurikar}, {Seifert}, {Sherbert}, {Sherwood-Taylor}, {Shih}, {Sick}, {Silbiger}, {Singanamalla}, {Singer}, {Sladen}, {Sooley}, {Sornarajah}, {Streicher}, {Teuben}, {Thomas}, {Tremblay}, {Turner}, {Terr{\'o}n}, {van Kerkwijk}, {de la Vega}, {Watkins}, {Weaver}, {Whitmore}, {Woillez}, {Zabalza}, \& {Astropy Contributors}}]{astropy:2018}
{Astropy Collaboration}, {Price-Whelan}, A.~M., {Sip{\H{o}}cz}, B.~M., {et~al.} 2018, \aj, 156, 123, \dodoi{10.3847/1538-3881/aabc4f}

\bibitem[{{Azzam} {et~al.}(2016){Azzam}, {Tennyson}, {Yurchenko}, \& {Naumenko}}]{azzam2016}
{Azzam}, A. A.~A., {Tennyson}, J., {Yurchenko}, S.~N., \& {Naumenko}, O.~V. 2016, \mnras, 460, 4063, \dodoi{10.1093/mnras/stw1133}

\bibitem[{{Bachmann} {et~al.}(2025){Bachmann}, {Kreidberg}, {Molli{\`e}re}, {Deming}, \& {Tsai}}]{bachman2025}
{Bachmann}, N., {Kreidberg}, L., {Molli{\`e}re}, P., {Deming}, D., \& {Tsai}, S.~M. 2025, arXiv e-prints, arXiv:2506.16232, \dodoi{10.48550/arXiv.2506.16232}

\bibitem[{{Baxter} {et~al.}(2020){Baxter}, {D{\'e}sert}, {Parmentier}, {Line}, {Fortney}, {Arcangeli}, {Bean}, {Todorov}, \& {Mansfield}}]{baxter2020}
{Baxter}, C., {D{\'e}sert}, J.-M., {Parmentier}, V., {et~al.} 2020, \aap, 639, A36, \dodoi{10.1051/0004-6361/201937394}

\bibitem[{{Beltz} {et~al.}(2021){Beltz}, {Rauscher}, {Brogi}, \& {Kempton}}]{beltz2021}
{Beltz}, H., {Rauscher}, E., {Brogi}, M., \& {Kempton}, E. M.~R. 2021, \aj, 161, 1, \dodoi{10.3847/1538-3881/abb67b}

\bibitem[{{Blain} {et~al.}(2024){Blain}, {Landman}, {Molli{\`e}re}, \& {Dittmann}}]{blain2024hd209}
{Blain}, D., {Landman}, R., {Molli{\`e}re}, P., \& {Dittmann}, J. 2024, \aap, 690, A63, \dodoi{10.1051/0004-6361/202450767}

\bibitem[{{Bonomo} {et~al.}(2017){Bonomo}, {Desidera}, {Benatti}, {Borsa}, {Crespi}, {Damasso}, {Lanza}, {Sozzetti}, {Lodato}, {Marzari}, {Boccato}, {Claudi}, {Cosentino}, {Covino}, {Gratton}, {Maggio}, {Micela}, {Molinari}, {Pagano}, {Piotto}, {Poretti}, {Smareglia}, {Affer}, {Biazzo}, {Bignamini}, {Esposito}, {Giacobbe}, {H{\'e}brard}, {Malavolta}, {Maldonado}, {Mancini}, {Martinez Fiorenzano}, {Masiero}, {Nascimbeni}, {Pedani}, {Rainer}, \& {Scandariato}}]{bonomo2017}
{Bonomo}, A.~S., {Desidera}, S., {Benatti}, S., {et~al.} 2017, \aap, 602, A107, \dodoi{10.1051/0004-6361/201629882}

\bibitem[{{Borysow}(2002)}]{borysow2002}
{Borysow}, A. 2002, \aap, 390, 779, \dodoi{10.1051/0004-6361:20020555}

\bibitem[{{Borysow} \& {Frommhold}(1989)}]{borysow1989b}
{Borysow}, A., \& {Frommhold}, L. 1989, \apj, 341, 549, \dodoi{10.1086/167515}

\bibitem[{{Borysow} {et~al.}(1989){Borysow}, {Frommhold}, \& {Moraldi}}]{borysow1989a}
{Borysow}, A., {Frommhold}, L., \& {Moraldi}, M. 1989, \apj, 336, 495, \dodoi{10.1086/167027}

\bibitem[{{Borysow} {et~al.}(2001){Borysow}, {Jorgensen}, \& {Fu}}]{borysow2001}
{Borysow}, A., {Jorgensen}, U.~G., \& {Fu}, Y. 2001, \jqsrt, 68, 235, \dodoi{10.1016/S0022-4073(00)00023-6}

\bibitem[{{Borysow} {et~al.}(1988){Borysow}, {Frommhold}, \& {Birnbaum}}]{borysow1988}
{Borysow}, J., {Frommhold}, L., \& {Birnbaum}, G. 1988, \apj, 326, 509, \dodoi{10.1086/166112}

\bibitem[{{Brogi} \& {Line}(2019)}]{brogi2019}
{Brogi}, M., \& {Line}, M.~R. 2019, \aj, 157, 114, \dodoi{10.3847/1538-3881/aaffd3}

\bibitem[{{Buchner} {et~al.}(2014){Buchner}, {Georgakakis}, {Nandra}, {Hsu}, {Rangel}, {Brightman}, {Merloni}, {Salvato}, {Donley}, \& {Kocevski}}]{buchner2014}
{Buchner}, J., {Georgakakis}, A., {Nandra}, K., {et~al.} 2014, \aap, 564, A125, \dodoi{10.1051/0004-6361/201322971}

\bibitem[{{Burrows} {et~al.}(2007){Burrows}, {Hubeny}, {Budaj}, {Knutson}, \& {Charbonneau}}]{burrows2007}
{Burrows}, A., {Hubeny}, I., {Budaj}, J., {Knutson}, H.~A., \& {Charbonneau}, D. 2007, \apjl, 668, L171, \dodoi{10.1086/522834}

\bibitem[{{Carvalho} \& {Johns-Krull}(2023)}]{carvalho2023}
{Carvalho}, A., \& {Johns-Krull}, C.~M. 2023, Research Notes of the American Astronomical Society, 7, 91, \dodoi{10.3847/2515-5172/acd37e}

\bibitem[{{Cheverall} {et~al.}(2023){Cheverall}, {Madhusudhan}, \& {Holmberg}}]{cheverall2023}
{Cheverall}, C.~J., {Madhusudhan}, N., \& {Holmberg}, M. 2023, \mnras, 522, 661, \dodoi{10.1093/mnras/stad648}

\bibitem[{{Cridland} {et~al.}(2019){Cridland}, {van Dishoeck}, {Alessi}, \& {Pudritz}}]{cridland2019}
{Cridland}, A.~J., {van Dishoeck}, E.~F., {Alessi}, M., \& {Pudritz}, R.~E. 2019, \aap, 632, A63, \dodoi{10.1051/0004-6361/201936105}

\bibitem[{{Cridland} {et~al.}(2020){Cridland}, {van Dishoeck}, {Alessi}, \& {Pudritz}}]{cridland2020}
---. 2020, \aap, 642, A229, \dodoi{10.1051/0004-6361/202038767}

\bibitem[{{Crossfield}(2023)}]{crossfield2023}
{Crossfield}, I. J.~M. 2023, \apjl, 952, L18, \dodoi{10.3847/2041-8213/ace35f}

\bibitem[{{Cutri} {et~al.}(2003){Cutri}, {Skrutskie}, {van Dyk}, {Beichman}, {Carpenter}, {Chester}, {Cambresy}, {Evans}, {Fowler}, {Gizis}, {Howard}, {Huchra}, {Jarrett}, {Kopan}, {Kirkpatrick}, {Light}, {Marsh}, {McCallon}, {Schneider}, {Stiening}, {Sykes}, {Weinberg}, {Wheaton}, {Wheelock}, \& {Zacarias}}]{cutri2003}
{Cutri}, R.~M., {Skrutskie}, M.~F., {van Dyk}, S., {et~al.} 2003, VizieR Online Data Catalog, II/246

\bibitem[{{Delorme} {et~al.}(2021){Delorme}, {Jovanovic}, {Echeverri}, {Mawet}, {Kent Wallace}, {Bartos}, {Cetre}, {Wizinowich}, {Ragland}, {Lilley}, {Wetherell}, {Doppmann}, {Wang}, {Morris}, {Ruffio}, {Martin}, {Fitzgerald}, {Ruane}, {Schofield}, {Suominen}, {Calvin}, {Wang}, {Magnone}, {Johnson}, {Sohn}, {L{\'o}pez}, {Bond}, {Pezzato}, {Sayson}, {Chun}, \& {Skemer}}]{kpic}
{Delorme}, J.-R., {Jovanovic}, N., {Echeverri}, D., {et~al.} 2021, Journal of Astronomical Telescopes, Instruments, and Systems, 7, 035006, \dodoi{10.1117/1.JATIS.7.3.035006}

\bibitem[{{Diamond-Lowe} {et~al.}(2014){Diamond-Lowe}, {Stevenson}, {Bean}, {Line}, \& {Fortney}}]{diamondlowe2014}
{Diamond-Lowe}, H., {Stevenson}, K.~B., {Bean}, J.~L., {Line}, M.~R., \& {Fortney}, J.~J. 2014, \apj, 796, 66, \dodoi{10.1088/0004-637X/796/1/66}

\bibitem[{{Dyrek} {et~al.}(2024){Dyrek}, {Min}, {Decin}, {Bouwman}, {Crouzet}, {Molli{\`e}re}, {Lagage}, {Konings}, {Tremblin}, {G{\"u}del}, {Pye}, {Waters}, {Henning}, {Vandenbussche}, {Ardevol Martinez}, {Argyriou}, {Ducrot}, {Heinke}, {van Looveren}, {Absil}, {Barrado}, {Baudoz}, {Boccaletti}, {Cossou}, {Coulais}, {Edwards}, {Gastaud}, {Glasse}, {Glauser}, {Greene}, {Kendrew}, {Krause}, {Lahuis}, {Mueller}, {Olofsson}, {Patapis}, {Rouan}, {Royer}, {Scheithauer}, {Waldmann}, {Whiteford}, {Colina}, {van Dishoeck}, {{\"O}stlin}, {Ray}, \& {Wright}}]{dyrek2024}
{Dyrek}, A., {Min}, M., {Decin}, L., {et~al.} 2024, \nat, 625, 51, \dodoi{10.1038/s41586-023-06849-0}

\bibitem[{Echeverri {et~al.}(2022)Echeverri, Jovanovic, Delorme, Xin, Schofield, Finnerty, Wang, Xuan, Mawet, Baker, Bartos, Bond, Bryan, Calvin, Cetre, Doppmann, Fitzgerald, Fucik, Horstman, Lopez, Martin, Martin, Mennesson, Morris, Nash, Pezzato, Porter, Ragland, Roberts, Ruane, Ruffio, Sappey, Serabyn, Skemer, Venenciano, Wallace, Wang, \& Wizinowich}]{echeverri2022}
Echeverri, D., Jovanovic, N., Delorme, J.-R., {et~al.} 2022, in Ground-based and Airborne Instrumentation for Astronomy IX, ed. C.~J. Evans, J.~J. Bryant, \& K.~Motohara, Vol. 12184, International Society for Optics and Photonics (SPIE), 121841W, \dodoi{10.1117/12.2630518}

\bibitem[{{Espinoza} {et~al.}(2017){Espinoza}, {Fortney}, {Miguel}, {Thorngren}, \& {Murray-Clay}}]{espinoza2017}
{Espinoza}, N., {Fortney}, J.~J., {Miguel}, Y., {Thorngren}, D., \& {Murray-Clay}, R. 2017, \apjl, 838, L9, \dodoi{10.3847/2041-8213/aa65ca}

\bibitem[{{Feroz} \& {Hobson}(2008)}]{feroz2008}
{Feroz}, F., \& {Hobson}, M.~P. 2008, \mnras, 384, 449, \dodoi{10.1111/j.1365-2966.2007.12353.x}

\bibitem[{{Feroz} {et~al.}(2009){Feroz}, {Hobson}, \& {Bridges}}]{feroz2009}
{Feroz}, F., {Hobson}, M.~P., \& {Bridges}, M. 2009, \mnras, 398, 1601, \dodoi{10.1111/j.1365-2966.2009.14548.x}

\bibitem[{{Feroz} {et~al.}(2019){Feroz}, {Hobson}, {Cameron}, \& {Pettitt}}]{feroz2019}
{Feroz}, F., {Hobson}, M.~P., {Cameron}, E., \& {Pettitt}, A.~N. 2019, The Open Journal of Astrophysics, 2, 10, \dodoi{10.21105/astro.1306.2144}

\bibitem[{Finnerty {et~al.}(2022)Finnerty, Schofield, Delorme, Sappey, Wang, Ruffio, Mawet, Fitzgerald, Jovanovic, Baker, Bartos, Bond, Bryan, Calvin, Cetre, Doppmann, Echeverri, Lopez, Martin, Morris, Pezzato, Ragland, Ruane, Skemer, Venenciano, Wallace, Wang, Wizinowich, \& Xuan}]{Finnerty2022}
Finnerty, L., Schofield, T., Delorme, J.-R., {et~al.} 2022, in Ground-based and Airborne Instrumentation for Astronomy IX, ed. C.~J. Evans, J.~J. Bryant, \& K.~Motohara, Vol. 12184, International Society for Optics and Photonics (SPIE), 121844Y, \dodoi{10.1117/12.2630276}

\bibitem[{{Finnerty} {et~al.}(2023){Finnerty}, {Schofield}, {Sappey}, {Xuan}, {Ruffio}, {Wang}, {Delorme}, {Blake}, {Buzard}, {Fitzgerald}, {Baker}, {Bartos}, {Bond}, {Calvin}, {Cetre}, {Doppmann}, {Echeverri}, {Jovanovic}, {Liberman}, {L{\'o}pez}, {Martin}, {Mawet}, {Morris}, {Pezzato}, {Phillips}, {Ragland}, {Skemer}, {Venenciano}, {Wallace}, {Wallack}, {Wang}, \& {Wizinowich}}]{finnerty2023}
{Finnerty}, L., {Schofield}, T., {Sappey}, B., {et~al.} 2023, \aj, 166, 31, \dodoi{10.3847/1538-3881/acda91}

\bibitem[{{Finnerty} {et~al.}(2024){Finnerty}, {Xuan}, {Xin}, {Liberman}, {Schofield}, {Fitzgerald}, {Agrawal}, {Baker}, {Bartos}, {Blake}, {Calvin}, {Cetre}, {Delorme}, {Doppmann}, {Echeverri}, {Hsu}, {Jovanovic}, {L{\'o}pez}, {Martin}, {Mawet}, {Morris}, {Pezzato}, {Ruffio}, {Sappey}, {Skemer}, {Venenciano}, {Wallace}, {Wallack}, {Wang}, \& {Wang}}]{finnerty2024}
{Finnerty}, L., {Xuan}, J.~W., {Xin}, Y., {et~al.} 2024, \aj, 167, 43, \dodoi{10.3847/1538-3881/ad1180}

\bibitem[{{Finnerty} {et~al.}(2025{\natexlab{a}}){Finnerty}, {Xin}, {Xuan}, {Inglis}, {Fitzgerald}, {Agrawal}, {Baker}, {Bartos}, {Blake}, {Calvin}, {Cetre}, {Delorme}, {Doppmann}, {Echeverri}, {Horstman}, {Hsu}, {Jovanovic}, {Liberman}, {L{\'o}pez}, {Martin}, {Mawet}, {Morris}, {Pezzato}, {Ruffio}, {Sappey}, {Schofield}, {Skemer}, {venenciano}, {Wallace}, {Wallack}, {Wang}, \& {Wang}}]{finnerty2025b}
{Finnerty}, L., {Xin}, Y., {Xuan}, J.~W., {et~al.} 2025{\natexlab{a}}, arXiv e-prints, arXiv:2503.01946, \dodoi{10.48550/arXiv.2503.01946}

\bibitem[{{Finnerty} {et~al.}(2025{\natexlab{b}}){Finnerty}, {Xin}, {Xuan}, {Inglis}, {Fitzgerald}, {Agrawal}, {Baker}, {Bartos}, {Blake}, {Calvin}, {Cetre}, {Delorme}, {Doppmann}, {Echeverri}, {Horstman}, {Hsu}, {Jovanovic}, {Liberman}, {L{\'o}pez}, {Martin}, {Mawet}, {Morris}, {Pezzato-Rovner}, {Ruffio}, {Sappey}, {Schofield}, {Skemer}, {Venenciano}, {Wallace}, {Wallack}, {Wang}, \& {Wang}}]{finnerty2025a}
---. 2025{\natexlab{b}}, AJ, 169, 94, \dodoi{10.3847/1538-3881/ada1d9}

\bibitem[{{Flowers} {et~al.}(2019){Flowers}, {Brogi}, {Rauscher}, {Kempton}, \& {Chiavassa}}]{flowers2019}
{Flowers}, E., {Brogi}, M., {Rauscher}, E., {Kempton}, E. M.~R., \& {Chiavassa}, A. 2019, \aj, 157, 209, \dodoi{10.3847/1538-3881/ab164c}

\bibitem[{Foreman-Mackey(2016)}]{corner}
Foreman-Mackey, D. 2016, The Journal of Open Source Software, 1, 24, \dodoi{10.21105/joss.00024}

\bibitem[{{Freedman} {et~al.}(2014){Freedman}, {Lustig-Yaeger}, {Fortney}, {Lupu}, {Marley}, \& {Lodders}}]{freedman2014}
{Freedman}, R.~S., {Lustig-Yaeger}, J., {Fortney}, J.~J., {et~al.} 2014, \apjs, 214, 25, \dodoi{10.1088/0067-0049/214/2/25}

\bibitem[{{Fu} {et~al.}(2024){Fu}, {Welbanks}, {Deming}, {Inglis}, {Zhang}, {Lothringer}, {Ih}, {Moses}, {Schlawin}, {Knutson}, {Henry}, {Greene}, {Sing}, {Savel}, {Kempton}, {Louie}, {Line}, \& {Nixon}}]{fu2024}
{Fu}, G., {Welbanks}, L., {Deming}, D., {et~al.} 2024, arXiv e-prints, arXiv:2407.06163, \dodoi{10.48550/arXiv.2407.06163}

\bibitem[{{Gaia Collaboration}(2020)}]{gaiaedr3}
{Gaia Collaboration}. 2020, VizieR Online Data Catalog, I/350

\bibitem[{{Gandhi} {et~al.}(2019){Gandhi}, {Madhusudhan}, {Hawker}, \& {Piette}}]{gandhi2019}
{Gandhi}, S., {Madhusudhan}, N., {Hawker}, G., \& {Piette}, A. 2019, \aj, 158, 228, \dodoi{10.3847/1538-3881/ab4efc}

\bibitem[{{Giacobbe} {et~al.}(2021){Giacobbe}, {Brogi}, {Gandhi}, {Cubillos}, {Bonomo}, {Sozzetti}, {Fossati}, {Guilluy}, {Carleo}, {Rainer}, {Harutyunyan}, {Borsa}, {Pino}, {Nascimbeni}, {Benatti}, {Biazzo}, {Bignamini}, {Chubb}, {Claudi}, {Cosentino}, {Covino}, {Damasso}, {Desidera}, {Fiorenzano}, {Ghedina}, {Lanza}, {Leto}, {Maggio}, {Malavolta}, {Maldonado}, {Micela}, {Molinari}, {Pagano}, {Pedani}, {Piotto}, {Poretti}, {Scandariato}, {Yurchenko}, {Fantinel}, {Galli}, {Lodi}, {Sanna}, \& {Tozzi}}]{giacobbe2021}
{Giacobbe}, P., {Brogi}, M., {Gandhi}, S., {et~al.} 2021, \nat, 592, 205, \dodoi{10.1038/s41586-021-03381-x}

\bibitem[{{Gordon} {et~al.}(2022){Gordon}, {Rothman}, {Hargreaves}, {Hashemi}, {Karlovets}, {Skinner}, {Conway}, {Hill}, {Kochanov}, {Tan}, {Wcis{\l}o}, {Finenko}, {Nelson}, {Bernath}, {Birk}, {Boudon}, {Campargue}, {Chance}, {Coustenis}, {Drouin}, {Flaud}, {Gamache}, {Hodges}, {Jacquemart}, {Mlawer}, {Nikitin}, {Perevalov}, {Rotger}, {Tennyson}, {Toon}, {Tran}, {Tyuterev}, {Adkins}, {Baker}, {Barbe}, {Can{\`e}}, {Cs{\'a}sz{\'a}r}, {Dudaryonok}, {Egorov}, {Fleisher}, {Fleurbaey}, {Foltynowicz}, {Furtenbacher}, {Harrison}, {Hartmann}, {Horneman}, {Huang}, {Karman}, {Karns}, {Kassi}, {Kleiner}, {Kofman}, {Kwabia-Tchana}, {Lavrentieva}, {Lee}, {Long}, {Lukashevskaya}, {Lyulin}, {Makhnev}, {Matt}, {Massie}, {Melosso}, {Mikhailenko}, {Mondelain}, {M{\"u}ller}, {Naumenko}, {Perrin}, {Polyansky}, {Raddaoui}, {Raston}, {Reed}, {Rey}, {Richard}, {T{\'o}bi{\'a}s}, {Sadiek}, {Schwenke}, {Starikova}, {Sung}, {Tamassia}, {Tashkun}, {Vander Auwera}, {Vasilenko}, {Vigasin}, {Villanueva}, {Vispoel}, {Wagner}, {Yachmenev}, \&
  {Yurchenko}}]{hitemp2020}
{Gordon}, I.~E., {Rothman}, L.~S., {Hargreaves}, R.~J., {et~al.} 2022, \jqsrt, 277, 107949, \dodoi{10.1016/j.jqsrt.2021.107949}

\bibitem[{{Guillot}(2010)}]{guillot2010}
{Guillot}, T. 2010, \aap, 520, A27, \dodoi{10.1051/0004-6361/200913396}

\bibitem[{{Hargreaves} {et~al.}(2020){Hargreaves}, {Gordon}, {Rey}, {Nikitin}, {Tyuterev}, {Kochanov}, \& {Rothman}}]{hargreaves2020}
{Hargreaves}, R.~J., {Gordon}, I.~E., {Rey}, M., {et~al.} 2020, \apjs, 247, 55, \dodoi{10.3847/1538-4365/ab7a1a}

\bibitem[{{Harris} {et~al.}(2006){Harris}, {Tennyson}, {Kaminsky}, {Pavlenko}, \& {Jones}}]{harris2006}
{Harris}, G.~J., {Tennyson}, J., {Kaminsky}, B.~M., {Pavlenko}, Y.~V., \& {Jones}, H.~R.~A. 2006, \mnras, 367, 400, \dodoi{10.1111/j.1365-2966.2005.09960.x}

\bibitem[{{Hawker} {et~al.}(2018){Hawker}, {Madhusudhan}, {Cabot}, \& {Gandhi}}]{hawker2018}
{Hawker}, G.~A., {Madhusudhan}, N., {Cabot}, S. H.~C., \& {Gandhi}, S. 2018, \apjl, 863, L11, \dodoi{10.3847/2041-8213/aac49d}

\bibitem[{{Henry} {et~al.}(2000){Henry}, {Marcy}, {Butler}, \& {Vogt}}]{henry2000}
{Henry}, G.~W., {Marcy}, G.~W., {Butler}, R.~P., \& {Vogt}, S.~S. 2000, \apjl, 529, L41, \dodoi{10.1086/312458}

\bibitem[{{Horstman} {et~al.}(2024){Horstman}, {Ruffio}, {Wang}, {Hsu}, {Baker}, {Finnerty}, {Xuan}, {Echeverri}, {Mawet}, {Blake}, {Bartos}, {Bond}, {Calvin}, {Cetre}, {Delorme}, {Doppmann}, {Fitzgerald}, {Jovanovic}, {Lopez}, {Martin}, {Morris}, {Pezzato}, {Ruane}, {Sappey}, {Schofield}, {Skemer}, {Venenciano}, {Wallace}, {Wang}, \& {Wizinowich}}]{horstman2024}
{Horstman}, K.~A., {Ruffio}, J.-B., {Wang}, J.~J., {et~al.} 2024, arXiv e-prints, arXiv:2408.10173, \dodoi{10.48550/arXiv.2408.10173}

\bibitem[{{Husser} {et~al.}(2013){Husser}, {Wende-von Berg}, {Dreizler}, {Homeier}, {Reiners}, {Barman}, \& {Hauschildt}}]{phoenix}
{Husser}, T.~O., {Wende-von Berg}, S., {Dreizler}, S., {et~al.} 2013, \aap, 553, A6, \dodoi{10.1051/0004-6361/201219058}

\bibitem[{{Inglis} {et~al.}(2024){Inglis}, {Batalha}, {Lewis}, {Kataria}, {Knutson}, {Kilpatrick}, {Gagnebin}, {Mukherjee}, {Pettyjohn}, {Crossfield}, {Foote}, {Grant}, {Henry}, {Lally}, {McKemmish}, {Sing}, {Wakeford}, {Zapata Trujillo}, \& {Zellem}}]{inglis2024}
{Inglis}, J., {Batalha}, N.~E., {Lewis}, N.~K., {et~al.} 2024, \apjl, 973, L41, \dodoi{10.3847/2041-8213/ad725e}

\bibitem[{{Jovanovic} {et~al.}(2025){Jovanovic}, {Echeverri}, {Delorme}, {Finnerty}, {Schofield}, {Wang}, {Xin}, {Xuan}, {Wallacee}, {Mawet}, {Sanghi}, {Baker}, {Bartos}, {Bond}, {Calvin}, {Cetre}, {Doppmann}, {Fitzgerald}, {Fucik}, {Gao}, {Ge}, {Guthery}, {Horstman}, {Hsud}, {Liberman}, {Leifer}, {Lilley}, {Lopez}, {Marin}, {Martin}, {Mennesson}, {Morris}, {Nash}, {Pezzato}, {Porter}, {Roberts}, {Ruane}, {Ruffio}, {Sappey}, {Serabyn}, {Shen}, {Skemer}, {Wang}, {Wetherell}, {Wizinowich}, {Salama}, {Chambouleyron}, {Jensen-Clem}, \& {Beichman}}]{kpicII}
{Jovanovic}, N., {Echeverri}, D., {Delorme}, J.-R., {et~al.} 2025, arXiv e-prints, arXiv:2502.01863, \dodoi{10.48550/arXiv.2502.01863}

\bibitem[{{Khorshid} {et~al.}(2022){Khorshid}, {Min}, {D{\'e}sert}, {Woitke}, \& {Dominik}}]{khorshid2022}
{Khorshid}, N., {Min}, M., {D{\'e}sert}, J.~M., {Woitke}, P., \& {Dominik}, C. 2022, \aap, 667, A147, \dodoi{10.1051/0004-6361/202141455}

\bibitem[{{Knutson} {et~al.}(2008){Knutson}, {Charbonneau}, {Allen}, {Burrows}, \& {Megeath}}]{knutson2008}
{Knutson}, H.~A., {Charbonneau}, D., {Allen}, L.~E., {Burrows}, A., \& {Megeath}, S.~T. 2008, \apj, 673, 526, \dodoi{10.1086/523894}

\bibitem[{{Kokori} {et~al.}(2023){Kokori}, {Tsiaras}, {Edwards}, {Jones}, {Pantelidou}, {Tinetti}, {Bewersdorff}, {Iliadou}, {Jongen}, {Lekkas}, \& et~al.}]{kokori23}
{Kokori}, A., {Tsiaras}, A., {Edwards}, B., {et~al.} 2023, \apjs, 265, 4, \dodoi{10.3847/1538-4365/ac9da4}

\bibitem[{{Lei} \& {Molli{\`e}re}(2024)}]{lei2024}
{Lei}, E., \& {Molli{\`e}re}, P. 2024, arXiv e-prints, arXiv:2410.21364, \dodoi{10.48550/arXiv.2410.21364}

\bibitem[{{Line} {et~al.}(2016){Line}, {Stevenson}, {Bean}, {Desert}, {Fortney}, {Kreidberg}, {Madhusudhan}, {Showman}, \& {Diamond-Lowe}}]{line2016}
{Line}, M.~R., {Stevenson}, K.~B., {Bean}, J., {et~al.} 2016, \aj, 152, 203, \dodoi{10.3847/0004-6256/152/6/203}

\bibitem[{{Line} {et~al.}(2021){Line}, {Brogi}, {Bean}, {Gandhi}, {Zalesky}, {Parmentier}, {Smith}, {Mace}, {Mansfield}, {Kempton}, {Fortney}, {Shkolnik}, {Patience}, {Rauscher}, {D{\'e}sert}, \& {Wardenier}}]{line2021}
{Line}, M.~R., {Brogi}, M., {Bean}, J.~L., {et~al.} 2021, \nat, 598, 580, \dodoi{10.1038/s41586-021-03912-6}

\bibitem[{{L{\'o}pez} {et~al.}(2020){L{\'o}pez}, {Hoffman}, {Doppmann}, {Fitzgerald}, {Johnson}, {Kassis}, {Lanclos}, {Lyke}, {Martin}, {McLean}, {Sohn}, \& {Weiss}}]{nirspecupgrade2}
{L{\'o}pez}, R.~A., {Hoffman}, E.~B., {Doppmann}, G., {et~al.} 2020, in Society of Photo-Optical Instrumentation Engineers (SPIE) Conference Series, Vol. 11447, Society of Photo-Optical Instrumentation Engineers (SPIE) Conference Series, 114476B, \dodoi{10.1117/12.2563075}

\bibitem[{{Lothringer} {et~al.}(2018){Lothringer}, {Barman}, \& {Koskinen}}]{lothringer2018}
{Lothringer}, J.~D., {Barman}, T., \& {Koskinen}, T. 2018, \apj, 866, 27, \dodoi{10.3847/1538-4357/aadd9e}

\bibitem[{{Lupu} {et~al.}(2014){Lupu}, {Zahnle}, {Marley}, {Schaefer}, {Fegley}, {Morley}, {Cahoy}, {Freedman}, \& {Fortney}}]{lupu2014}
{Lupu}, R.~E., {Zahnle}, K., {Marley}, M.~S., {et~al.} 2014, \apj, 784, 27, \dodoi{10.1088/0004-637X/784/1/27}

\bibitem[{{Madhusudhan} {et~al.}(2017){Madhusudhan}, {Bitsch}, {Johansen}, \& {Eriksson}}]{madhusudhan2017}
{Madhusudhan}, N., {Bitsch}, B., {Johansen}, A., \& {Eriksson}, L. 2017, \mnras, 469, 4102, \dodoi{10.1093/mnras/stx1139}

\bibitem[{{Martin} {et~al.}(2018){Martin}, {Fitzgerald}, {McLean}, {Doppmann}, {Kassis}, {Aliado}, {Canfield}, {Johnson}, {Kress}, {Lanclos}, {Magnone}, {Sohn}, {Wang}, \& {Weiss}}]{nirspecupgrade}
{Martin}, E.~C., {Fitzgerald}, M.~P., {McLean}, I.~S., {et~al.} 2018, in Society of Photo-Optical Instrumentation Engineers (SPIE) Conference Series, Vol. 10702, Ground-based and Airborne Instrumentation for Astronomy VII, ed. C.~J. {Evans}, L.~{Simard}, \& H.~{Takami}, 107020A, \dodoi{10.1117/12.2312266}

\bibitem[{{McLean} {et~al.}(1998){McLean}, {Becklin}, {Bendiksen}, {Brims}, {Canfield}, {Figer}, {Graham}, {Hare}, {Lacayanga}, {Larkin}, {Larson}, {Levenson}, {Magnone}, {Teplitz}, \& {Wong}}]{nirspec}
{McLean}, I.~S., {Becklin}, E.~E., {Bendiksen}, O., {et~al.} 1998, in Society of Photo-Optical Instrumentation Engineers (SPIE) Conference Series, Vol. 3354, Infrared Astronomical Instrumentation, ed. A.~M. {Fowler}, 566--578, \dodoi{10.1117/12.317283}

\bibitem[{{Molli{\`e}re} {et~al.}(2019){Molli{\`e}re}, {Wardenier}, {van Boekel}, {Henning}, {Molaverdikhani}, \& {Snellen}}]{prt:2019}
{Molli{\`e}re}, P., {Wardenier}, J.~P., {van Boekel}, R., {et~al.} 2019, \aap, 627, A67, \dodoi{10.1051/0004-6361/201935470}

\bibitem[{{Molli{\`e}re} {et~al.}(2020){Molli{\`e}re}, {Stolker}, {Lacour}, {Otten}, {Shangguan}, {Charnay}, {Molyarova}, {Nowak}, {Henning}, {Marleau}, {Semenov}, {van Dishoeck}, {Eisenhauer}, {Garcia}, {Garcia Lopez}, {Girard}, {Greenbaum}, {Hinkley}, {Kervella}, {Kreidberg}, {Maire}, {Nasedkin}, {Pueyo}, {Snellen}, {Vigan}, {Wang}, {de Zeeuw}, \& {Zurlo}}]{prt:2020}
{Molli{\`e}re}, P., {Stolker}, T., {Lacour}, S., {et~al.} 2020, \aap, 640, A131, \dodoi{10.1051/0004-6361/202038325}

\bibitem[{{NASA Exoplanet Science Institute}(2020)}]{10.26133/nea12}
{NASA Exoplanet Science Institute}. 2020, Planetary Systems Table,  IPAC, \dodoi{10.26133/NEA12}

\bibitem[{{NASA Exoplanet Science Institute}(2022)}]{10.26134/ExoFOP5}
---. 2022, Exoplanet Follow-up Observing Program Web Service,  IPAC, \dodoi{10.26134/EXOFOP5}

\bibitem[{Nasedkin {et~al.}(2024)Nasedkin, Mollière, \& Blain}]{Nasedkin2024}
Nasedkin, E., Mollière, P., \& Blain, D. 2024, Journal of Open Source Software, 9, 5875, \dodoi{10.21105/joss.05875}

\bibitem[{{{\"O}berg} {et~al.}(2011){{\"O}berg}, {Murray-Clay}, \& {Bergin}}]{oberg2011}
{{\"O}berg}, K.~I., {Murray-Clay}, R., \& {Bergin}, E.~A. 2011, \apjl, 743, L16, \dodoi{10.1088/2041-8205/743/1/L16}

\bibitem[{{Partridge} \& {Schwenke}(1997)}]{partridge1997}
{Partridge}, H., \& {Schwenke}, D.~W. 1997, \jcp, 106, 4618, \dodoi{10.1063/1.473987}

\bibitem[{{Polyansky} {et~al.}(2018){Polyansky}, {Kyuberis}, {Zobov}, {Tennyson}, {Yurchenko}, \& {Lodi}}]{polyansky2018}
{Polyansky}, O.~L., {Kyuberis}, A.~A., {Zobov}, N.~F., {et~al.} 2018, \mnras, 480, 2597, \dodoi{10.1093/mnras/sty1877}

\bibitem[{{Powell} {et~al.}(2024){Powell}, {Feinstein}, {Lee}, {Zhang}, {Tsai}, {Taylor}, {Kirk}, {Bell}, {Barstow}, {Gao}, {Bean}, {Blecic}, {Chubb}, {Crossfield}, {Jordan}, {Kitzmann}, {Moran}, {Morello}, {Moses}, {Welbanks}, {Yang}, {Zhang}, {Ahrer}, {Bello-Arufe}, {Brande}, {Casewell}, {Crouzet}, {Cubillos}, {Demory}, {Dyrek}, {Flagg}, {Hu}, {Inglis}, {Jones}, {Kreidberg}, {L{\'o}pez-Morales}, {Lagage}, {Meier Vald{\'e}s}, {Miguel}, {Parmentier}, {Piette}, {Rackham}, {Radica}, {Redfield}, {Stevenson}, {Wakeford}, {Aggarwal}, {Alam}, {Batalha}, {Batalha}, {Benneke}, {Berta-Thompson}, {Brady}, {Caceres}, {Carter}, {D{\'e}sert}, {Harrington}, {Iro}, {Line}, {Lothringer}, {MacDonald}, {Mancini}, {Molaverdikhani}, {Mukherjee}, {Nixon}, {Oza}, {Palle}, {Rustamkulov}, {Sing}, {Steinrueck}, {Venot}, {Wheatley}, \& {Yurchenko}}]{powell2024}
{Powell}, D., {Feinstein}, A.~D., {Lee}, E. K.~H., {et~al.} 2024, \nat, 626, 979, \dodoi{10.1038/s41586-024-07040-9}

\bibitem[{{Rosenthal} {et~al.}(2021){Rosenthal}, {Fulton}, {Hirsch}, {Isaacson}, {Howard}, {Dedrick}, {Sherstyuk}, {Blunt}, {Petigura}, {Knutson}, {Behmard}, {Chontos}, {Crepp}, {Crossfield}, {Dalba}, {Fischer}, {Henry}, {Kane}, {Kosiarek}, {Marcy}, {Rubenzahl}, {Weiss}, \& {Wright}}]{rosenthal2021}
{Rosenthal}, L.~J., {Fulton}, B.~J., {Hirsch}, L.~A., {et~al.} 2021, \apjs, 255, 8, \dodoi{10.3847/1538-4365/abe23c}

\bibitem[{{Roth} {et~al.}(2024){Roth}, {Parmentier}, \& {Hammond}}]{roth2024}
{Roth}, A., {Parmentier}, V., \& {Hammond}, M. 2024, \mnras, 531, 1056, \dodoi{10.1093/mnras/stae984}

\bibitem[{{Rothman} {et~al.}(2010){Rothman}, {Gordon}, {Barber}, {Dothe}, {Gamache}, {Goldman}, {Perevalov}, {Tashkun}, \& {Tennyson}}]{hitemp2010}
{Rothman}, L.~S., {Gordon}, I.~E., {Barber}, R.~J., {et~al.} 2010, \jqsrt, 111, 2139, \dodoi{10.1016/j.jqsrt.2010.05.001}

\bibitem[{{Rothman} {et~al.}(2013){Rothman}, {Gordon}, {Babikov}, {Barbe}, {Chris Benner}, {Bernath}, {Birk}, {Bizzocchi}, {Boudon}, {Brown}, {Campargue}, {Chance}, {Cohen}, {Coudert}, {Devi}, {Drouin}, {Fayt}, {Flaud}, {Gamache}, {Harrison}, {Hartmann}, {Hill}, {Hodges}, {Jacquemart}, {Jolly}, {Lamouroux}, {Le Roy}, {Li}, {Long}, {Lyulin}, {Mackie}, {Massie}, {Mikhailenko}, {M{\"u}ller}, {Naumenko}, {Nikitin}, {Orphal}, {Perevalov}, {Perrin}, {Polovtseva}, {Richard}, {Smith}, {Starikova}, {Sung}, {Tashkun}, {Tennyson}, {Toon}, {Tyuterev}, \& {Wagner}}]{rothman2013}
{Rothman}, L.~S., {Gordon}, I.~E., {Babikov}, Y., {et~al.} 2013, \jqsrt, 130, 4, \dodoi{10.1016/j.jqsrt.2013.07.002}

\bibitem[{{Rustamkulov} {et~al.}(2023){Rustamkulov}, {Sing}, {Mukherjee}, {May}, {Kirk}, {Schlawin}, {Line}, {Piaulet}, {Carter}, {Batalha}, {Goyal}, {L{\'o}pez-Morales}, {Lothringer}, {MacDonald}, {Moran}, {Stevenson}, {Wakeford}, {Espinoza}, {Bean}, {Batalha}, {Benneke}, {Berta-Thompson}, {Crossfield}, {Gao}, {Kreidberg}, {Powell}, {Cubillos}, {Gibson}, {Leconte}, {Molaverdikhani}, {Nikolov}, {Parmentier}, {Roy}, {Taylor}, {Turner}, {Wheatley}, {Aggarwal}, {Ahrer}, {Alam}, {Alderson}, {Allen}, {Banerjee}, {Barat}, {Barrado}, {Barstow}, {Bell}, {Blecic}, {Brande}, {Casewell}, {Changeat}, {Chubb}, {Crouzet}, {Daylan}, {Decin}, {D{\'e}sert}, {Mikal-Evans}, {Feinstein}, {Flagg}, {Fortney}, {Harrington}, {Heng}, {Hong}, {Hu}, {Iro}, {Kataria}, {Kempton}, {Krick}, {Lendl}, {Lillo-Box}, {Louca}, {Lustig-Yaeger}, {Mancini}, {Mansfield}, {Mayne}, {Miguel}, {Morello}, {Ohno}, {Palle}, {Petit dit de la Roche}, {Rackham}, {Radica}, {Ramos-Rosado}, {Redfield}, {Rogers}, {Shkolnik}, {Southworth}, {Teske}, {Tremblin},
  {Tucker}, {Venot}, {Waalkes}, {Welbanks}, {Zhang}, \& {Zieba}}]{rustamkulov2023}
{Rustamkulov}, Z., {Sing}, D.~K., {Mukherjee}, S., {et~al.} 2023, \nat, 614, 659, \dodoi{10.1038/s41586-022-05677-y}

\bibitem[{{Savel} {et~al.}(2025){Savel}, {Bedell}, {Kempton}, {Smith}, {Bean}, {Zhao}, {Wong}, {Sanchez}, \& {Line}}]{savel2025}
{Savel}, A.~B., {Bedell}, M., {Kempton}, E. M.~R., {et~al.} 2025, \aj, 169, 135, \dodoi{10.3847/1538-3881/ada27e}

\bibitem[{{Schwarz} {et~al.}(2015){Schwarz}, {Brogi}, {de Kok}, {Birkby}, \& {Snellen}}]{schwarz2015}
{Schwarz}, H., {Brogi}, M., {de Kok}, R., {Birkby}, J., \& {Snellen}, I. 2015, \aap, 576, A111, \dodoi{10.1051/0004-6361/201425170}

\bibitem[{{Showman} {et~al.}(2009){Showman}, {Fortney}, {Lian}, {Marley}, {Freedman}, {Knutson}, \& {Charbonneau}}]{showman2009}
{Showman}, A.~P., {Fortney}, J.~J., {Lian}, Y., {et~al.} 2009, \apj, 699, 564, \dodoi{10.1088/0004-637X/699/1/564}

\bibitem[{{Snellen} {et~al.}(2010){Snellen}, {de Kok}, {de Mooij}, \& {Albrecht}}]{snellen2010}
{Snellen}, I. A.~G., {de Kok}, R.~J., {de Mooij}, E. J.~W., \& {Albrecht}, S. 2010, \nat, 465, 1049, \dodoi{10.1038/nature09111}

\bibitem[{{Tsai} {et~al.}(2023){Tsai}, {Lee}, {Powell}, {Gao}, {Zhang}, {Moses}, {H{\'e}brard}, {Venot}, {Parmentier}, {Jordan}, {Hu}, {Alam}, {Alderson}, {Batalha}, {Bean}, {Benneke}, {Bierson}, {Brady}, {Carone}, {Carter}, {Chubb}, {Inglis}, {Leconte}, {Line}, {L{\'o}pez-Morales}, {Miguel}, {Molaverdikhani}, {Rustamkulov}, {Sing}, {Stevenson}, {Wakeford}, {Yang}, {Aggarwal}, {Baeyens}, {Barat}, {de Val-Borro}, {Daylan}, {Fortney}, {France}, {Goyal}, {Grant}, {Kirk}, {Kreidberg}, {Louca}, {Moran}, {Mukherjee}, {Nasedkin}, {Ohno}, {Rackham}, {Redfield}, {Taylor}, {Tremblin}, {Visscher}, {Wallack}, {Welbanks}, {Youngblood}, {Ahrer}, {Batalha}, {Behr}, {Berta-Thompson}, {Blecic}, {Casewell}, {Crossfield}, {Crouzet}, {Cubillos}, {Decin}, {D{\'e}sert}, {Feinstein}, {Gibson}, {Harrington}, {Heng}, {Henning}, {Kempton}, {Krick}, {Lagage}, {Lendl}, {Lothringer}, {Mansfield}, {Mayne}, {Mikal-Evans}, {Palle}, {Schlawin}, {Shorttle}, {Wheatley}, \& {Yurchenko}}]{tsai2023}
{Tsai}, S.-M., {Lee}, E. K.~H., {Powell}, D., {et~al.} 2023, \nat, 617, 483, \dodoi{10.1038/s41586-023-05902-2}

\bibitem[{{Turrini} {et~al.}(2021){Turrini}, {Schisano}, {Fonte}, {Molinari}, {Politi}, {Fedele}, {Pani{\'c}}, {Kama}, {Changeat}, \& {Tinetti}}]{turrini2021}
{Turrini}, D., {Schisano}, E., {Fonte}, S., {et~al.} 2021, \apj, 909, 40, \dodoi{10.3847/1538-4357/abd6e5}

\bibitem[{{Xue} {et~al.}(2024){Xue}, {Bean}, {Zhang}, {Welbanks}, {Lunine}, \& {August}}]{xue2024}
{Xue}, Q., {Bean}, J.~L., {Zhang}, M., {et~al.} 2024, \apjl, 963, L5, \dodoi{10.3847/2041-8213/ad2682}

\bibitem[{{Yurchenko} {et~al.}(2018){Yurchenko}, {Al-Refaie}, \& {Tennyson}}]{exocross2018}
{Yurchenko}, S.~N., {Al-Refaie}, A.~F., \& {Tennyson}, J. 2018, \aap, 614, A131, \dodoi{10.1051/0004-6361/201732531}

\bibitem[{{Yurchenko} {et~al.}(2011){Yurchenko}, {Barber}, \& {Tennyson}}]{yurchenko2011}
{Yurchenko}, S.~N., {Barber}, R.~J., \& {Tennyson}, J. 2011, \mnras, 413, 1828, \dodoi{10.1111/j.1365-2966.2011.18261.x}

\end{thebibliography}
\bibliographystyle{aasjournal}



\end{CJK*}
\end{document}